\documentclass{jfm}

\usepackage[T1]{fontenc}
\usepackage[english]{babel}
\usepackage[utf8x]{inputenc}

\usepackage{graphicx}
\usepackage{subcaption}

\usepackage{lipsum}
\usepackage{titletoc}

\usepackage{amstext}
\usepackage{amsmath,amssymb}

\usepackage{array}
\usepackage{color}
\usepackage{colortbl}

\usepackage[titletoc]{appendix}
\usepackage{esint}

\usepackage{array,multirow}

\usepackage{hyperref}
\hypersetup{
    colorlinks=true,
    linkcolor=red,
    citecolor=blue,
    filecolor=cyan,
}

\newcommand\vect[1]{ \mathbf{#1} }

\newcommand\tder[2]{\frac{\textrm{d} #1}{\textrm{d} #2}}

\newcommand\avg[1]{\left\langle #1 \right\rangle}
\newcommand\norm[1]{\left\Vert #1 \right\Vert}
\newcommand\of[1]{\left( #1 \right)}

\newcommand\orderof[1]{\textit{O}\left( #1 \right)}

\newcommand{\de}{\text{\textrm{d}}}
\newcommand{\St}{\text{\textit{St}}}

\graphicspath{{./figures/}}

\shorttitle{Multiscale fluid--particle thermal interaction}
\shortauthor{M. Carbone, A. D. Bragg and M. Iovieno}

\title{Multiscale fluid--particle thermal interaction in isotropic turbulence}

\author{M. Carbone\aff{1},
  A. D. Bragg\aff{2}
  \corresp{\email{andrew.bragg@duke.edu}}
 \and M. Iovieno\aff{1}}

\affiliation{\aff{1}Dipartimento di Ingegneria Meccanica e Aerospaziale,
Politecnico di Torino, Corso Duca degli Abruzzi 24, 10129 Torino, Italy
\aff{2}Department of Civil and Environmental Engineering, Duke University, Durham, NC 27708,
USA}

\begin{document}

\maketitle

\begin{abstract}
We use direct numerical simulations to investigate the interaction between the temperature field of a fluid and the temperature of small particles suspended in the flow, employing both one and two-way thermal coupling, in a statistically stationary, isotropic turbulent flow. Using statistical analysis, we investigate this variegated interaction at the different scales of the flow. We find that the variance of the fluid temperature gradients decreases as the thermal response time of the suspended particles is increased. The probability density function (PDF) of the fluid temperature gradients scales with its variance, while the PDF of the rate of change of the particle temperature, whose variance is associated with the thermal dissipation due to the particles, does not scale in such a self-similar way. The modification of the fluid temperature field due to the particles is examined by computing the particle concentration and particle heat fluxes conditioned on the magnitude of the local fluid temperature gradient. These statistics highlight that the particles cluster on the fluid temperature fronts, and the important role played by the alignments of the particle velocity and the local fluid temperature gradient.
The temperature structure functions, which characterize the temperature fluctuations across the scales of the flow, clearly show that the fluctuations of the fluid temperature increments are monotonically suppressed in the two-way coupled regime as the particle thermal response time is increased.
Thermal caustics dominate the particle temperature increments at small scales, that is, particles that come into contact are likely to have very large differences in their temperature. This is caused by the nonlocal thermal dynamics of the particles, and the scaling exponents of the inertial particle temperature structure functions in the dissipation range  reveal very strong multifractal behavior.
Further insight is provided by the PDFs of the two-point temperature increments and by the flux of temperature increments across the scales. All together, these results reveal a number of non-trivial effects, with a number of important practical consequences.
\end{abstract}


\section{Introduction}

The interaction between inertial particles and scalar fields in turbulent flows plays a central role in many natural problems, ranging from cloud microphysics \citep{Pruppacher2010,Grabowski2013} to the interactions between plankton and nutrients \citep{DeLillo2014}, and dust particle flows in accretion disks \citep{Takeuchi2002}. In engineered systems, applications involve chemical reactors and combustion chambers, and more recently, microdispersed colloidal fluids where the enhanced thermal conductivity due to particle aggregations can give rise to non-trivial thermal behavior \citep{Prasher2006, Momenifar2015}, and which can be used in cooling devices for electronic equipment exposed to large heat fluxes \citep{Das2006}. 

In this work, we focus on the heat exchange between advected inertial particles and the fluid phase in a turbulent flow, with a parametric emphasis relevant to understanding particle-scalar interactions in cloud microphysics. Understanding the droplet growth in clouds requires to characterize the interaction between water droplets and the humidity and temperature fields. A major problem is to understand how the interaction between turbulence, heat exchange, condensational processes, and collisions can produce the rapid growth of water droplets that leads to rain initiation \citep{Pruppacher2010,Grabowski2013}.
While the study of the transport of scalar fields and particles in turbulent flows are well established research areas in both theoretical and applied fluid dynamics \citep{Kraichnan1994,Taylor1922}, the characterization of the interaction between scalars and particles in turbulent flows is a relatively new topic \citep{Bec2014}, since the problem is hard to handle analytically, requires sophisticated experimental techniques, and is computationally demanding.

When temperature differences inside the fluid are sufficiently small, the temperature field behaves almost like a passive scalar, that is, the fluid temperature is advected and diffused by the fluid motion but has negligible dynamical effect on the flow. Even in this regime, the statistical properties of the passive scalar field are significantly different from those of the underlying velocity field that advects it. Different regimes take place according to the Reynolds number and the ratio between momentum and scalar diffusivities \citep{Shraiman2000,Warhaft2000,Watanabe2004}. 

Experiments, numerical simulations and analytical models show that a passive scalar field is always more intermittent than the velocity field, and passive scalars in turbulence are characterized by strong anomalous scaling \citep{Holzer1994}. This is due to the formation of ramp--cliff structures in the scalar field \citep{Celani2000,Watanabe2004}: large regions in which the scalar field is almost constant are separated by thin regions in which the scalar abruptly changes. The regions in which the scalar mildly changes are referred to as Lagrangian coherent structures. The thin regions with large scalar gradient, where the diffusion of the scalar takes place, are referred to as fronts. It has been shown that the large scale forcing influences the passive scalar statistics at small scales \citep{Gotoh2015}. In particular, a mean scalar gradient forcing preserves universality of the statistics while a large scale Gaussian forcing does not. However, the ramp-cliff structure was observed with different types of forcing, implying that this structure is universal to scalar fields in turbulence \citep{Watanabe2004,Bec2014}. Moreover, recent measurements of atmospheric turbulence have shown that external boundary conditions, such as the magnitude and sign of the sensible heat flux, have a significant impact on the fluid temperature dynamics within the inertial range, while for the same scales the fluid velocity increments are essentially independent of these large-scale conditions \citep{zorzetto18}.

When a turbulent flow is seeded with inertial particles, the particles can sample the surrounding flow in a non-uniform and correlated manner \citep{Toschi2009}. Particle inertia in a turbulent flow is measured through the Stokes number $\St \equiv\tau_p/\tau_\eta$, which compares the particle response time to the Kolmogorov time scale. A striking feature of inertial particle motion in turbulent flows is that they spontaneously cluster even in incompressible flows \citep{maxey87,wang93,Bec2007,Ireland2016}. This clustering can take place across a wide range of scales \citep{Bec2007,Bragg2015b,Ireland2016}, and the small-scale clustering is maximum when $\St=\orderof{1}$. A variety of mechanisms has been proposed to explain this phenomena: when $\St\ll1$ the clustering is caused by particles being centrifuged out of regions of strong rotation \citep{maxey87,chun05}, while for $\St\geq \orderof{1}$, a non-local mechanism generates the clustering, whose effect is related to the particles memory of its interaction with the flow along its path-history \citep{gustavsson11b,gustavsson16,bragg14b,bragg2015a,Bragg2015b}. Note that recent results on the clustering of settling inertial particles in turbulence have corroborated this picture, showing that strong clustering can occur even in a parameter regime where the centrifuge effect cannot be invoked as the explanation for the clustering, but is caused by a non-local mechanism \citep{ireland16b}. 

When particles have finite thermal inertia, they will not be in thermal equilibrium with the fluid temperature field, and this can give rise to non-trivial thermal coupling between the fluid and particles in a turbulent flow. A thermal response time $\tau_\theta$ can be defined so that the particle thermal inertia is parameterized by the thermal Stokes number $\St_\theta \equiv \tau_\theta/\tau_\eta$ \citep{Zaichik2009}. Since both the fluid temperature and particle phase-space dynamics depend upon the fluid velocity field, there can exist non-trivial correlations between the fluid and particle temperatures even in the absence of thermal coupling. Indeed, it was show by \cite{Bec2014} that inertial particles preferentially cluster on the fronts of the scalar field. Associated with this is that the particles preferentially sample the fluid temperature field, and when combined with the strong intermittency of temperature fields in turbulent flows, that can cause particles to experience very large temperature fluctuations along their trajectories. 

Several works have considered aspects of the fluid-particle temperature coupling using numerical simulations. For example, \cite{Zonta2008} investigated a particle-laden channel flow, with a view to modeling the modification of heat transfer in micro--dispersed fluids. They considered both momentum and temperature two--way coupling and observed that, depending on the particle inertia, the heat flow at the wall can increase or decrease. \cite{Kuerten2011} considered a similar set-up with larger dispersed particles, and they observed a stronger modification of the fluid temperature statistics due to the particles. \cite{Zamansky2014,Zamansky2016} considered turbulence induced by buoyancy, where the buoyancy was generated by heated particles. They observed that the resulting flow is driven by thermal plumes produced by the particles. As the particle inertia was increased, the inhomogeneity and the effect of the coupling were enhanced in agreement with the fact that inertial particles tend to cluster on the scalar fronts. 
\cite{Kumar2014} examined how the spatial distribution of droplets is affected by large scale inhomogeneities in the fluid temperature and supersaturation fields, considering the transition between homogeneous and inhomogeneous mixing.
A similar flow configuration was also investigated by \cite{Gotzfried2017}.

Each of these studies was primarily focused on the effect of the inertial particles on the large-scale statistics of the fluid temperature field. However, the results of \cite{Bec2014} imply that the effects of fluid-particle thermal coupling could be strong at the small scales, owing to the fact that they cluster on the fronts of the temperature field. Moreover, there is a need to understand and characterize the multiscale thermal properties of the particles themselves. In order to address these issues, we have conducted direct numerical simulations (DNS) to investigate the interaction between the scalar temperature field and the temperature of inertial particles suspended in the fluid, with one and two-way thermal coupling, in statistically stationary, isotropic turbulence. Using statistical analysis, we probe the multiscale aspects of the problem and consider the particular ways that the inertial particles contribute to the properties of the fluid temperature field in the two-way coupled regime.

The paper is organized as follows. In section \ref{sec:simul} we present the physical model used in the DNS, and present the parameters in the system. In section \ref{sec:diss} the statistics of the fluid temperature and time derivative of the particle temperature are considered, which allow us to quantify the contributions to the thermal dissipation in the system from the fluid and particles. In section \ref{sec:fluct} we consider the statistics of the fluid and particle temperature. In section \ref{sec:FATT} we consider the heat flux due to the particle motion conditioned on the local fluid temperature gradients in order to obtain insight into the details of the thermal coupling. In section \ref{sec:SF} we consider the structure functions of the fluid and particle temperature increments, along with their scaling exponents. In section \ref{sec:2pPDF} we consider the probability density functions (PDFs) of the fluid and particle temperature increments, along with the PDFs of the fluxes of the fluid and particle temperature increments across the scales of the flow. Finally, concluding remarks are given in section \ref{sec:concl}.

\section{The physical model}
\label{sec:simul}

In this section we present the governing equations of the physical model which will be solved numerically to simulate the thermal coupling and behavior of a particle-laden turbulent flow.
\subsection{Fluid phase}

We consider a statistically stationary, homogeneous and isotropic turbulent flow, governed by the incompressible Navier-Stokes equations. The turbulent velocity field advects the fluid temperature field (assumed a passive scalar), together with the inertial particles. In this study, we account for two-way thermal coupling between the fluid and particles, but only one-way momentum coupling. Therefore, the governing equations for the fluid phase are
\begin{eqnarray}
\bnabla\bcdot\vect{u} &=& 0 \label{NScontinuity},\\
\partial_t \vect{u} + \vect{u}\bcdot\bnabla\vect{u}  &=& -\frac{1}{\rho_0}\bnabla p + \nu\nabla^2 \vect{u} + \vect{f} \label{NSmomentum},\\
\partial_t T + \vect{u}\bcdot\bnabla T &=& \kappa \nabla^2 T - C_T + f_T \label{NSscalar}.
\end{eqnarray}
Here $\vect{u}\left(\vect{x},t\right)$ is the velocity of the fluid, $p\left(\vect{x},t\right)$ is the pressure, $\rho_0$ is the density of the fluid, $\nu$ is its kinematic viscosity, $T\left(\vect{x},t\right)$ is the temperature of the fluid and $\kappa$ is the thermal diffusivity. The ratio between the the momentum diffusivity $\nu$ and the thermal diffusivity $\kappa$ defines the Prandtl number $\Pran\equiv\nu/\kappa$. In this work, we consider $\Pran=1$, leaving further exploration of its effect on the system to future work. The $\vect{f}$ and $f_T$ terms in equations \eqref{NSmomentum} and \eqref{NSscalar} represent the external forcing, and $C_T$ is the thermal feedback of the particles on the fluid temperature field, that is, the heat exchanged per unit time and unit volume between the fluid and particles at position $\vect{x}$.

When the forcing is confined to sufficiently large scales, it is assumed that the details of the forcing do not influence the small-scale dynamics. Previous experimental evidence seems to confirm this \citep{Sreenivasan1996}, leading to a universal behaviour of the small-scales. However, recent studies \citep{Gotoh2015} pointed out that this hypothesis of universality is partially violated by the advected scalar fields, whose inertial range statistics exhibit sensitivity to the details of the imposed forcing. Since we aim to characterize temperature and temperature gradient fluctuations in the dissipation range for different inertia of the suspended particles, we employ a forcing that imposes the same total dissipation rate for all the simulations. This produces results which can be meaningfully compared for different parameters of the suspended particles, since the response of the system to the same injected thermal power can be examined. Therefore, we employ a large scale forcing which imposes the average dissipation rate \citep{Kumar2014}, that is
\begin{equation}
\hat{\vect{f}}(\vect{k}) = \varepsilon \frac{\hat{\vect{u}}(\vect{k})}
{\sum_{\vect{k}\in\mathcal{K}_f} \left\Vert \hat{\vect{u}}(\vect{k}) \right\Vert^2 },
\quad
\hat{f}_T(\vect{k}) = \chi\frac{\hat{T}(\vect{k})}{\sum_{\vect{k}\in\mathcal{K}_f} \left\vert \hat{T}(\vect{k}) \right\vert^2 },
\end{equation}
in the wavenumber space (a hat indicates the Fourier transform  and $\vect{k}$ is the wavenumber). Here $\mathcal{K}_f$ is the set of forced wavenumbers while $\varepsilon$ and $\chi$ are the imposed dissipation rates of velocity and temperature variance, respectively. Since both the velocity and temperature statistics at large scales tend to be close to Gaussian, this forcing behaves similarly to a random Gaussian forcing. The value of the parameters relative to the fluid flow, employed in the simulations are in table \ref{tab:flow}.

\begin{table}
\centering
\begin{tabular}{l c c}
Kinematic viscosity &	$\nu$ & 0.005 \\
Prandtl number & $\Pran$ & 1 \\
Velocity fluctuations dissipation rate & $\varepsilon$ &	0.27 \\
Temperature fluctuations dissipation rate & $\chi$ &	$0.1$ \\
Kolmogorov time scale	& $\tau_\eta$& 0.136\\
Kolmogorov length scale	& $\eta$ & 0.0261\\
Taylor micro-scale	& $\lambda$ & 0.498\\
Integral length scale & $\ell$	& 1.4\\
Root mean square velocity & $u'$ & 0.88 \\
Kolmogorov velocity scale	& $u_\eta$& 0.192\\
Small scale temperature  & $T_\eta$ &	$0.117$ \\
Taylor Reynolds number	& $\Rey_\lambda$ & 88\\
Integral scale Reynolds number	& $\Rey_l$ & 244\\
Forced wavenumber	& $k_f$ & $\sqrt{2}$\\
Number of Fourier modes	& $N$  & $128$ (3/2)\\
Resolution	& $N\eta/2$ & 1.67
\end{tabular}
\caption{Flow parameters in dimensionless code units. The characteristic parameters of the fluid flow are defined from its energy spectrum $E\of{k}\equiv\int_{\norm{\vect{k}}=k}\left\Vert \vect{\hat{u}}\of{\vect{k}} \right\Vert^2 \de\vect{k}/2$. The dissipation rate of turbulent kinetic energy is: $\varepsilon \equiv2\nu\int k^2 E\of{k} \de k$. The Kolmogorov length $\eta\equiv\left(\nu^3/\varepsilon\right)^{1/4}$, time scale $\tau_\eta\equiv\left(\nu/\varepsilon\right)^{1/2}$ and velocity scale $u_\eta \equiv \eta/\tau_\eta$. The Taylor micro-scale is: $\lambda\equiv u'/\sqrt{\langle\left|\bnabla \vect{u}\right|^2\rangle}$. The root mean square velocity is $u'\equiv\sqrt{(2/3)\int E\of{k} \de k}$ and the integral length scale $\ell \equiv \left.\upi\middle/\left(2u'^2\right)\right. \int E\of{k}/k \de k$. Similarly, the spectrum, root mean square value and dissipation rate of the scalar field are: $E_T\of{k}\equiv\int_{\norm{\vect{k}}=k}\left\vert \hat{T}\of{\vect{k}} \right\vert^2 \de\vect{k}/2$, $T' \equiv\sqrt{(1/2)\int  E_{T}\of{k} \de k}$, $\chi \equiv 2\kappa \int  k^2 E_T\of{k} \de k$. The small scale temperature is determined by the viscosity and dissipation rate: $T_\eta\equiv\sqrt{\chi\tau_\eta}$. Since the Prandtl number is unitary the small scales of the scalar and the velocity field are of the same order.}
\label{tab:flow}
\end{table}
\subsection{Particle phase}

We consider rigid, point-like particles which are heavy with respect to the fluid, and small with respect to any scale of the flow. In particular, the particle density $\rho_p$ is much larger than the fluid density $\rho_p\gg\rho_0$, and the particle radius $r_p$ is much smaller than the Kolmogorov length scale $r_p \ll \eta$. With these assumptions (and neglecting gravity) the particle acceleration is described by the Stokes drag law. Analogously, the rate of change of the particle temperature is described by Newton's law for the heat conduction
\begin{eqnarray}
\tder{\vect{x}_p}{t} &\equiv& \vect{v}_p,
\label{eq:partx}\\
\tder{\vect{v}_p}{t} &=& \frac{ \vect{u} \left(\vect{x}_{p},t \right) - \vect{v}_p }{\tau_p},
\label{eq:partv}\\
\tder{\theta_p}{t}   &=& \frac{ T \left( \vect{x}_{p} , t \right) - \theta_{p}  }{\tau_\theta}.
\label{eq:parttheta}
\end{eqnarray}
Here $\tau_p \equiv\left.2\rho_pr_p^2\middle/\left(9\rho_0\nu\right)\right.$ is the particle momentum response time, $\tau_\theta \equiv \left. \rho_p c_p r_p^2\middle/\left(3\rho_0 c_0\kappa\right)\right.$ is the particle thermal response time, $c_p$ is the particle heat capacity, and $c_0$ is the fluid heat capacity at constant pressure. The Stokes number is defined as $\St \equiv \left.\tau_p\middle/\tau_\eta\right.$ ,and the thermal Stokes number is defined as $\St_\theta \equiv \left.\tau_\theta\middle/\tau_\eta\right.$, where $\tau_\eta$ is the Kolmogorov time scale.  

Our simulations focus on a dilute suspension regime with particle volume fraction $\phi=4\times 10^{-4}$. While this volume fraction is large enough for two-way momentum coupling between the particles and fluid to be important \citep[e.g.][]{elghobashi91}, we ignore this in the present study. The motivation is that including both two-way momentum and two-way thermal coupling introduces too many competing effects that would compound a thorough understanding of the problem. In this study we therefore ignore momentum coupling, but account for two-way thermal coupling, and in a follow up study we will include the effects of two-way momentum coupling.

We consider nine values of $\St_\theta$ and three values of $\St$ in order to explore the behavior of the system over a range of parameter values. Since we are accounting for thermal coupling, each combination of $\St_\theta$ and $\St$ must be simulated separately, and when combined with the large number of particles in the flow domain, the set of simulations require considerable computational resources. Therefore, in the present study we restrict attention to $\Rey_\lambda =88$, but future explorations should consider larger $\Rey_\lambda$ in order to explore the behavior when there exists a well-defined inertial range in the flow. 

In order to obtain deeper insight into the role of the two-way thermal coupling, we perform simulations with (denoted by S1) and without (denoted by S2) the thermal coupling. The particle parameters employed in the simulations are in table \ref{tab:particles}.

\subsection{Thermal coupling}

In the two-way thermal coupling regime, the thermal energy contained in the fluid is finite with respect to the thermal energy of the particles, therefore, when heat flows from the fluid to the particle the fluid loses thermal energy at the particle position. Due to the point-mass approximation, the feedback from the particles on the fluid temperature field is a superposition of Dirac delta functions, centered on the particles. Hence the coupling term in equation \eqref{NSscalar} is given by
\begin{equation}
C_T \left(\vect{x},t\right) = \frac{4}{3}\upi\frac{\rho_p}{\rho_0}\frac{c_p}{c_0} r_p^3 \sum_{p=1}^{N_P}\tder{\theta_p}{t}\delta\left(\vect{x} - \vect{x}_p\right).
\label{eq:CT}
\end{equation}

\begin{table}
\centering
\begin{tabular}{l c c}
Particle phase volume fraction &$\phi$ & $0.0004$  \\
Particle to fluid density ratio &$\rho_p/\rho_0$ & $1000$  \\
Particle back reaction & $C_T$ & S1: included; S2: neglected.\\
Stokes number &$\St$ & $0.5$; $1$; $3$. \\
Thermal Stokes number & $\St_\theta$ & $0.2$; $0.5$; $1$; $1.5$; $2$; $3$; $4$; $5$; $6$.\\
Number of particles & $N_P$ & $12500992$; $4419584$; $847872$.
\end{tabular}
\caption{Particles parameters in dimensionless code units. The Stokes number is $\St\equiv\tau_p/\tau_\eta$ and thermal Stokes number $\St_\theta\equiv\tau_\theta/\tau_\eta$ and the particle response times are defined in the text. In the simulations, $\St_\theta$ is varied by varying the particle heat capacity. The different combinations of $\St$ and $\St_\theta$ are simulated including the two-way thermal thermal coupling (simulations S1) and neglecting it (simulations S2).}
\label{tab:particles}
\end{table}


\begin{figure}
\centering
\includegraphics[width=\textwidth]{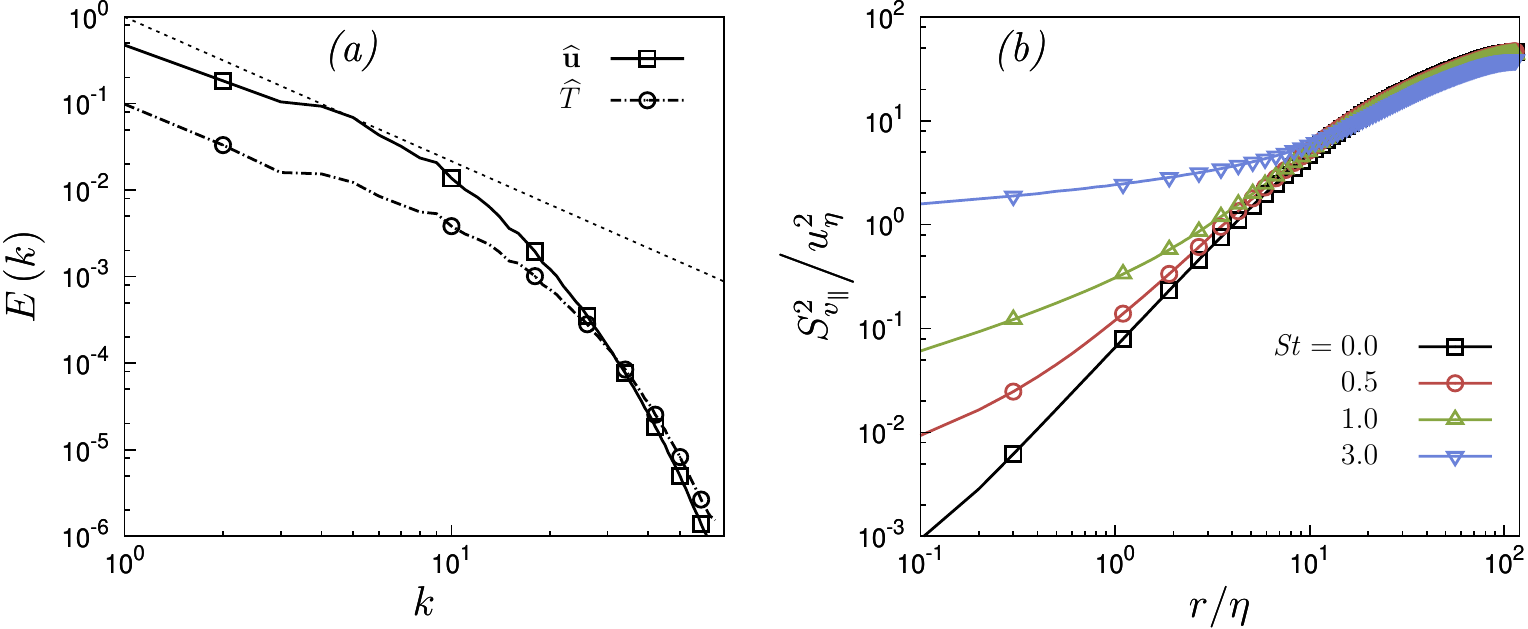}
\caption{(\textit{a}) Three-dimensional energy spectrum of the fluid velocity field (open squares) and temperature field (open circles). The temperature field is computed without any feedback from the particles on the fluid flow (simulations S2). \protect  (\textit{b}) Second order longitudinal structure functions of the particle velocity for various Stokes numbers.}
\label{fig:flowstat}
\end{figure}

\subsection{Numerical method}

We perform direct numerical simulation of incompressible, statistically steady and isotropic turbulence on a tri-periodic cubic domain. Equations \eqref{NScontinuity}, \eqref{NSmomentum}, and \eqref{NSscalar} are solved by means of the pseudo-spectral Fourier method for the spatial discretization. The $3/2$ rule is employed for dealiasing \citep{Canuto1988}, so that the maximum resolved wavenumber is $k_{\textrm{max}}=N/2$. The required Fourier transforms are executed in parallel using the P3DFFT library \citep{Pekurovsky2012}. Forcing is applied to a single scale, that is to all wavevectors satisfying $\left\Vert \vect{k} \right\Vert^2 = k_f$, with $k_f=2$, and the equations for the fluid velocity and temperature Fourier coefficients are evolved in time by means of a second order Runge-Kutta exponential integrator \citep{Hochbruck2010}. This method has been preferred to the standard integrating factor because of its higher accuracy and, above all, because of its consistency. Indeed, in order to obtain an accurate representation of small scale temperature fluctuations, it is critical that the numerical solution conserves thermal energy. The same time integration scheme is used to solve particle equations \eqref{eq:partx}, \eqref{eq:partv} and \eqref{eq:parttheta}, thus providing overall consistency, since the system formed by fluid and particles is evolved in time as a whole.

The fluid velocity and temperature are interpolated at the particle position by means of fourth order B-spline interpolation. The interpolation is implemented as a backward Non Uniform Fourier Transform with B-spline basis: the fluid field is projected onto the B-spline basis in Fourier space through a deconvolution, than transformed into the physical space by means of a inverse Fast Fourier Transform (FFT). A convolution provides the interpolated field at particle position \citep{Beylkin1995}. Since B-splines have a compact support in physical space and deconvolution in Fourier space reduces to a division, this provide an efficient way to obtain high order interpolation.
This guarantees smooth and accurate interpolation and its efficient implementation is suitable for pseudo-spectral methods \citep{vanHinsberg2012}.
Moreover, the same method is used to obtain the spectral representation of the coupling term \eqref{eq:CT}.
The coupling term has to be projected on the Cartesian grid used to represent the fields. This is performed by means of the forward Nonuniform Fast Fourier Transform (NUFFT) with B-spline basis \citep{Beylkin1995}. Briefly, the algorithm works as follows. The convolution of the distribution $C_T\of{\vect{x},t}$ with the B-spline polynomial basis $B\of{\vect{x}}$ is computed in physical space, so that it can be effectively represented on the Cartesian grid
\begin{equation}
\widetilde{C}\left(\vect{x},t \right) = \int C_T\of{\vect{x},t} B\of{\vect{x}-\vect{y}}\de\vect{y}.
\end{equation}
Then, the regularized field $\widetilde{C}$ is transformed to Fourier space and the convolution with the B-spline basis is efficiently removed:
\begin{equation}
\widehat{{C}}\of{\vect{k},t} = \frac{\widehat{\widetilde{C}}\of{\vect{k},t}}{\widehat{B}\of{\vect{k}}}.
\end{equation}
This algorithm allows an efficient and accurate spectral representation of the particle back-reaction \citep{Carbone2018}. Indeed, the NUFFT satisfies the constraints for interpolation schemes \citep{Sundaram1996}: the backward and forward transformations are symmetric and the non locality, introduced in physical space due to the convolution, is removed in Fourier space. For these reasons this technique is preferred to shape regularization functions, \citep{Maxey1997}.

\section{Characterization of the thermal dissipation rate}
\label{sec:diss}
In the flow under consideration, the total dissipation rate of the temperature field $\chi$ is constant due to the forcing term $f_T$.  The total dissipation has a contribution from the fluid and particle phases and is given by \citep{Sundaram1996}
\begin{equation}
\chi = \kappa \avg{ \left\Vert \bnabla T \right\Vert^2 } +  \frac{\phi}{\tau_\theta}\frac{\rho_p c_p}{\rho_0 c_0} \avg{\left( T \of{\vect{x}_p,t} - \theta_p \right)^2}.\label{eq:dissbal0}
\end{equation}
We indicate with $\chi_f$ the dissipation due to the fluid temperature gradient and with $\chi_p$ the dissipation due to the particles, the two terms in the right hand side of equation \eqref{eq:dissbal0}, so that $\chi=\chi_f+\chi_p$. Note that both contributions to the dissipation rate are proportional to the kinematic thermal conductivity of the fluid since $\tau_\theta\propto 1/\kappa$, and hence both the dissipation mechanisms are due to molecular diffusivity. 

A characteristic length of the dissipation due to the particles can be defined as
\begin{equation}
\eta_p \equiv \frac{r_p}{\sqrt{3 \phi }}\label{eq:disslen}
\end{equation}
and using this, the balance of the dissipation of the temperature fluctuations can be written as
\begin{equation}
\chi = \kappa \left[ \avg{ \left\Vert \bnabla T \right\Vert^2 } + \avg{\left(\frac{ T \of{\vect{x}_p,t} - \theta_p}{\eta_p}\right)^2 } \right].\label{eq:dissbal}
\end{equation}
In these simulations the volume fraction $\phi$ is constant, so the characteristic length of the dissipation due to the particles is proportional to the particle radius. 

The portion of temperature fluctuations dissipated by the two different mechanisms depends on the statistics of the differences between the particle and local fluid temperatures. In the limit $\St_\theta\to 0$ we have $T \of{\vect{x}_p,t}=\theta_p$, such that all of the dissipation is associated with the fluid. In the general case, the statistics of $T \of{\vect{x}_p,t}-\theta_p$ depend not only on $\St_\theta$, but also implicitly upon $\St$, with the statistics of $T \of{\vect{x}_p,t}$ depending on the spatial clustering of the particles. This coupling between the particle momentum and temperature dynamics can lead to non-trivial effects of particle inertia on $\chi_p$.

\subsection{Thermal dissipation due to the temperature gradients}
Since the flow is isotropic, $\chi_f$ is given by
\begin{equation}
\chi_f = 3\kappa\avg{\of{\partial_x T}^2}
\end{equation}
We consider fixed Reynolds number and $\Pran=1$, thus $\kappa$ is the same in all the presented simulations, and so $\langle(\partial_x T)^2\rangle$ fully characterizes $\chi_f$. Moreover, given the expected structure of the field $\partial_x T$, it is instructive to consider its full Probability Density Function (PDF), in addition to its moments in order to know how different regions of the flow contribute to the average dissipation rate $\chi_f$.
\begin{figure}
\centering
\includegraphics[width=\textwidth]{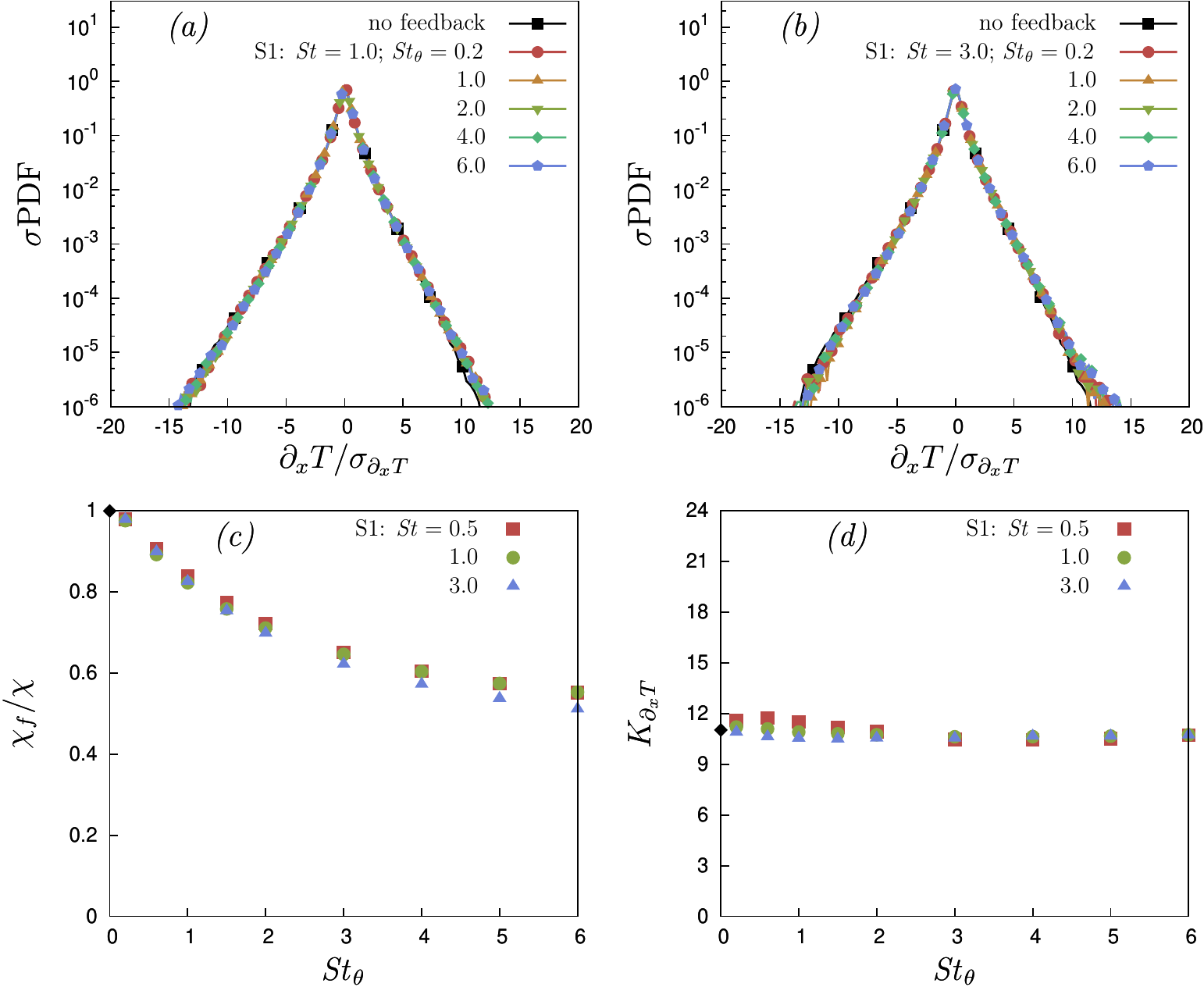}
\caption{PDF of the fluid temperature gradient $\partial_x T$ from simulations S1, for $\St=1$ (\textit{a}) and $\St=3$ (\textit{b}), and for various $\St_\theta$. (\textit{c}) Dissipation rate $\chi_f$ of the fluid temperature fluctuations, for different $\St$ as a function of $\St_\theta$. (\textit{d}) Kurtosis of the fluid temperature gradient PDF.}
\label{fig:PDFgrT}
\end{figure}

Figures \ref{fig:PDFgrT}(a-b) show the normalized PDFs of $\partial_x T$ for $\St=1$ and $\St=3$ respectively, and for various $\St_\theta$, where the PDFs are normalized using the standard deviation of the distribution, $\sigma_{\partial_x T}$. The distribution is almost symmetric and it displays elongated exponential tails. The largest temperature gradients exceed the standard deviation by an order of magnitude \citep{Overholt1996}. Remarkably, the shape of the PDF shows a very weak dependence on $\St$ and $\St_\theta$, such that the PDF shape scales with $\sigma_{\partial_x T}$.

The variance of the fluid temperature gradient is proportional to the actual dissipation rate of the temperature fluctuation (the proportionality factor being $3\kappa$, fixed in our simulations). In contrast to the PDF shape, the  suspended particles have a strong impact on $\chi_f$, as shown in figure \ref{fig:PDFgrT}. As $\St_\theta$ is increased, $\chi_f$ decreases. However, this is mainly due to the fact that as $\St_\theta$ is increased, $\chi_p$ increases, and so $\chi_f$ must decrease since $\chi=\chi_f+\chi_p$ is fixed. The influence of the Stokes number on $\chi_p$ is very small in the range of parameters considered.

The kurtosis of the fluid temperature gradients is shown in figure \ref{fig:PDFgrT}(d), as a function of $\St_\theta$ and for various $\St$. The kurtosis is approximately constant, and much larger than the value of a Gaussian distribution. The behavior of the kurtosis confirms that the fluid temperature gradient PDF is approximately self-similar.

\subsection{Thermal dissipation due to the particle dynamics}

The dissipation rate due to the particles, $\chi_p$, depends on the difference between the particle temperature and the fluid temperature at the particle position
\begin{equation}
\chi_p = \kappa\avg{\left(\frac{ T \of{\vect{x}_p,t} - \theta_p}{\eta_p}\right)^2 }
\end{equation}
For notational simplicity, we define $\varphi_p \equiv \left.\left(T \of{\vect{x}_p,t} - \theta_p\right)\middle/\eta_p\right.$. When $\varphi_p$ is normalized by its standard deviation, we can relate this to the rate of change of the particle temperature using equation \eqref{eq:parttheta}
\begin{equation}
\frac{ \dot{\theta}_p }{ \sigma_{ \dot{\theta}_p } } = \frac{\varphi_p}{\sigma_{\varphi_p}}.
\end{equation}

\begin{figure}
\centering
\includegraphics[width=\textwidth]{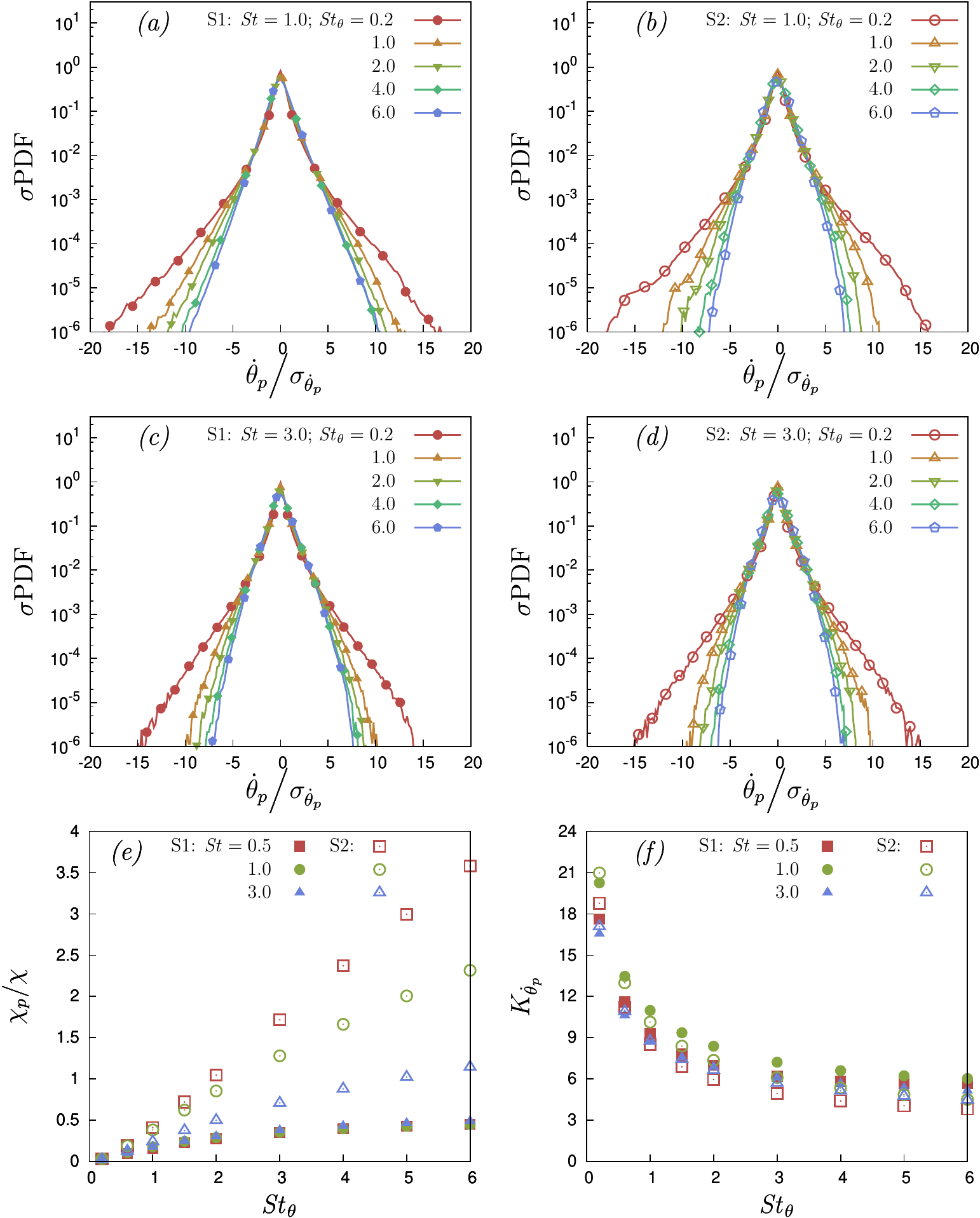}
\caption{PDF of $\dot{\theta}_p$ for $\St=1$ (\textit{a-b}) and $\St=3$ (\textit{c-d}), and for various $\St_\theta$. Plots (\textit{a-c}) are from simulations S1, in which the two-way thermal coupling is considered, while plots (\textit{b-d}) are from simulations S2, in which the two-way coupling is neglected. (\textit{e}) Dissipation rate $\chi_p$ of the temperature fluctuations due to the particles, for different $\St$ as a function of $\St_\theta$. (\textit{f}) Kurtosis of the PDF of $\dot{\theta}_p$.}
\label{fig:PDFdeT}
\end{figure}

The normalized PDF of $\dot{\theta}_p$ for $\St=1$ and $\St=3$, and for various $\St_\theta$ is shown in figure \ref{fig:PDFdeT}. Figure \ref{fig:PDFdeT}(a) shows the normalized PDF of $\dot{\theta}_p$, for $\St=1$ for the set of simulations S1, in which the two-way thermal coupling is taken to account. Figure \ref{fig:PDFdeT}(b) shows the corresponding results for simulations S2, in which the two-way thermal coupling is neglected. The normalized PDF of $\dot{\theta}_p$ for $\St=3$, with and without the two-way thermal coupling, is shown in figures \ref{fig:PDFdeT}(c-d). 

\begin{figure}
\centering
\includegraphics[width=\textwidth]{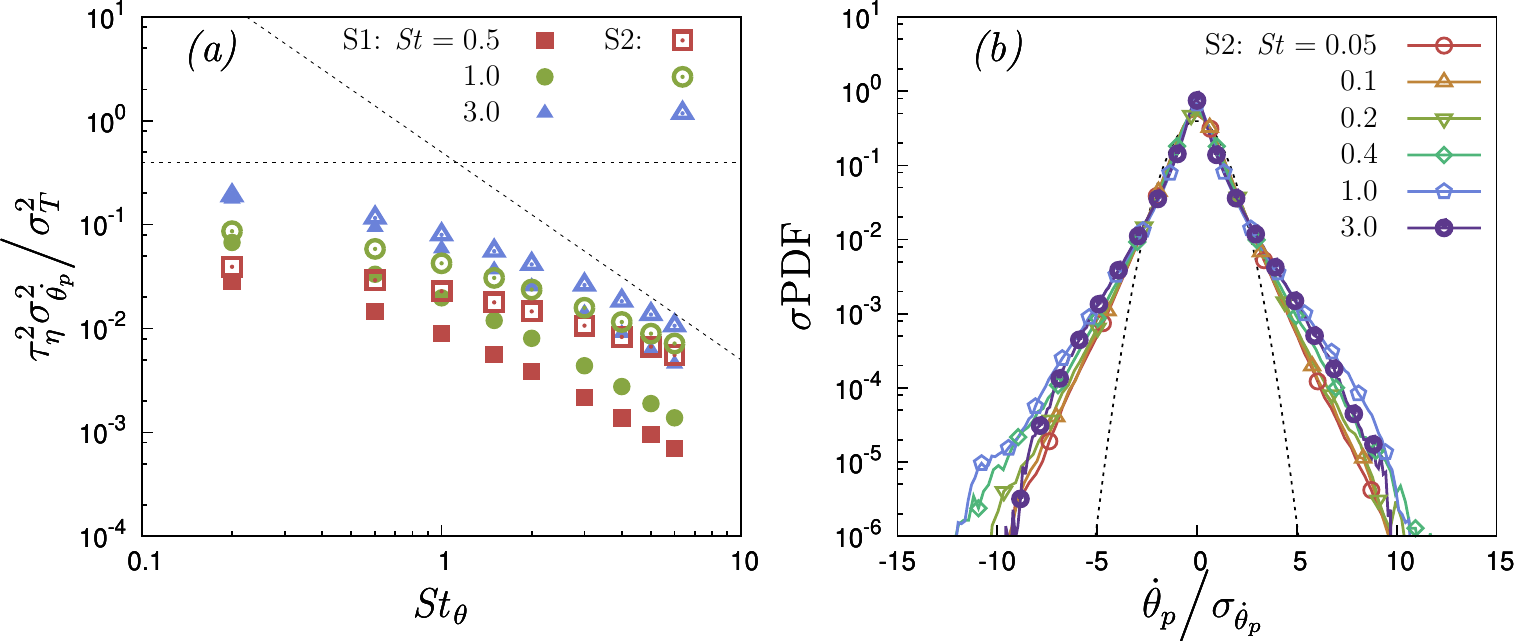}
\caption{(a) Variance of the particle temperature rate of change as a function of the thermal Stokes number for different Stokes numbers. The dotted lines represent the expected asymptotic behaviour for $\St_\theta\ll1$ and $\St_\theta\gg1$. (b) Normalized PDF of the particle temperature rate of change,  $\dot{\theta_p}$ at $\St_\theta=1$ for various Stokes number, $\St<0.5$. The dotted line shows a Gaussian PDF for reference. Results obtained neglecting the particle thermal feedback.}
\label{fig:compare}
\end{figure}

In contrast to the fluid temperature gradient PDFs, the shape of the PDF of $\dot{\theta}_p$ is not self-similar with respect to its variance. As $\St_\theta$ is increased, the normalized PDF becomes narrower. This is due to the fact that as $\St_\theta$ is increased, the particles respond more slowly to changes in the fluid temperature field, analogous to the ``filtering'' effect for inertial particle velocities in turbulence \citep{salazar12a,Ireland2016}. The PDF shapes are mildly affected by $\St$, and for larger $\St_\theta$, extreme fluid temperature-particle temperature differences are suppressed when the two-way thermal coupling is neglected.

The variance of $\dot{\theta}_p$ is proportional to the particle dissipation rate $\chi_p$, and the results for this are shown in figure \ref{fig:PDFdeT}(e), for various $\St$ and $\St_\theta$, and for simulations S1 and S2. The results show that as $\St_\theta$ is increased, $\chi_p$ increases. This is mainly because as $\St_\theta$ is increased, the thermal memory of the particle increases, and the particle temperature depends strongly on its encounter with the fluid temperature field along its trajectory history for times up to $\orderof{\tau_\theta}$ in the past. As a result, the particle temperature can differ strongly from the local fluid temperature. The results also show that $\chi_p$ is dramatically suppressed when two-way thermal coupling is accounted for. One reason for this is that as shown earlier, two-way thermal coupling leads to a suppression in the fluid temperature gradients. As these gradients are suppressed, the fluid temperature along the particle trajectory history differs less from the local fluid temperature than it would have in the absence of two-way thermal coupling, and as a result $\chi_p$ is decreased.

The results for kurtosis of $\dot{\theta}_p$, as a function of $\St_\theta$ and for various $\St$ are shown in figure \ref{fig:PDFdeT}(f). The results show that the kurtosis decreases with increasing $\St_\theta$. This is mainly due to the filtering effect mentioned earlier, wherein as $\St_\theta$ is increased, the particles are less able to respond to rapid fluctuations in the fluid temperature along their trajectory. Further, the kurtosis is typically larger when the two-way thermal coupling is taken into account (simulations S1), and is maximum for $\St=1$. This is due to the particle clustering on the fronts of the fluid temperature field, as will be discussed in section \ref{sec:FATT}.

Our results for the PDF of $\dot{\theta}_p$ and its moments differ somewhat from those in \cite{Bec2014}. This is in part due to the difference in the forcing methods employed by \cite{Bec2014} and that in our study. The solution of \eqref{eq:parttheta} may be written as \citep{Bec2014}
\begin{equation}
\avg{\dot{\theta_p}^2} = \frac{1}{2\tau_\theta^3}\int_0^\infty \avg{\Big( \delta_t T_p(t)\Big)^2}\exp\of{-\frac{t}{\tau_\theta}}\de t,
\label{eq:solthetalag}
\end{equation}
where $\delta_t T_p(t)\equiv T\of{\vect{x}_p\of{t},t}-T\of{\vect{x}_p\of{0},0}$.

In the regime $\St_\theta\ll1$, the exponential in \eqref{eq:solthetalag} decays very fast in time so that the main contribution to the integral comes from $\delta_t T_p$ for infinitesimal $t$, with $\delta_t T_p \sim t^n$ for $t\to 0$. Substituting $\delta_t T_p \sim t^n$ into \eqref{eq:solthetalag} we obtain the leading order behavior\
\begin{equation}
\avg{\dot{\theta_p}^2} \sim \frac{1}{2\tau_\theta^3}\int_0^\infty t^{2n} \exp\of{-\frac{t}{\tau_\theta}}\de t \sim \St_\theta^{2n-2},\;\St_\theta\ll 1.
\end{equation}
\cite{Bec2014} used a white in time forcing for the fluid scalar field, giving $n=1/2$, and yielding $\langle\dot{\theta_p}^2\rangle \sim \St_\theta^{-1}$ for  $\St_\theta\ll1$. However, the forcing scheme that we have employed generates a field $T(\vect{x},t)$ that evolves smoothly in time, so $n=1$ and $\langle\dot{\theta_p}^2\rangle\sim$ constant for $\St_\theta\ll1$. 

For $\St_\theta\gg 1$, the integral in \eqref{eq:solthetalag} is dominated by uncorrelated temperature increments, $\delta_t T \sim t^0$, such that $\langle\dot{\theta_p}^2\rangle \sim \St_\theta^{-2}$. The comparison between figure \ref{fig:compare}(a) and figure 5 of \cite{Bec2014} highlights the different asymptotic behavior of $\sigma_{\dot{\theta_p}}^2\equiv \langle\dot{\theta_p}^2\rangle$ for $\St_\theta\ll 1$, but the same behavior $\langle\dot{\theta_p}^2\rangle \sim \St_\theta^{-2}$ for $\St_\theta\gg 1$. Further, as expected, our DNS data approaches these asymptotic regimes for both the cases with and without two-way thermal coupling.

Another difference is that in the results of \cite{Bec2014}, the tails of the PDFs of $\dot{\theta}_p$ for $\St_\theta=1$ become heavier as $\St$ is increased, whereas our results in figure \ref{fig:PDFdeT} show that while the kurtosis of these PDFs increases from $\St=0.5$ to $\St=1$, it then decreases from $\St=1$ to $\St=3$. In order to examine this further, we performed simulations (without two-way thermal coupling) for $\St_\theta=1$ and $\St\leq 0.4$. The results are shown in figure \ref{fig:compare}(b), and in this regime we do in fact observe that the tails of the PDFs of $\dot{\theta}_p$ become increasingly wider as $\St$ is increased. Taken together with the results in figure \ref{fig:PDFdeT}, this implies that in our simulations, the tails of the PDFs of $\dot{\theta}_p$ become increasingly wider as $\St$ is increased until $\St\approx 1$, where this behavior then saturates, and upon further increase of $\St$ the tails start to narrow. This non-monotonic behavior is due to the particle clustering in the fronts of the temperature field, which is strongest for $\St\approx 1$ (see \S\ref{sec:FATT}). While the results in \cite{Bec2014} over the range $\St\leq  3.7$ do not show the tails of the PDFs of $\dot{\theta}_p$ becoming narrower, their results clearly show that the widening of the tails saturates (see inset of figure 5 in \cite{Bec2014}). It is possible that if they had considered larger $\St$, they would have also began to observe a narrowing of the tails as $\St$ was further increased. Possible reasons why the widening of the tails saturates at a lower value of $\St$ in our DNS than it does in theirs include is the effect of Reynolds number ($\Rey_\lambda=315$ in their DNS, whereas in our DNS $\Rey_\lambda=88$), and differences in the scalar forcing method.

\section{Characterization of the temperature fluctuations}
\label{sec:fluct}

This section consists of a short overview of the one-point temperature statistics. Note that due to the large scale forcing used in the DNS, the one-point statistics of the flow are affected by the forcing method employed.
\subsection{Fluid temperature fluctuations}

\begin{figure}
\centering
\includegraphics[width=\textwidth]{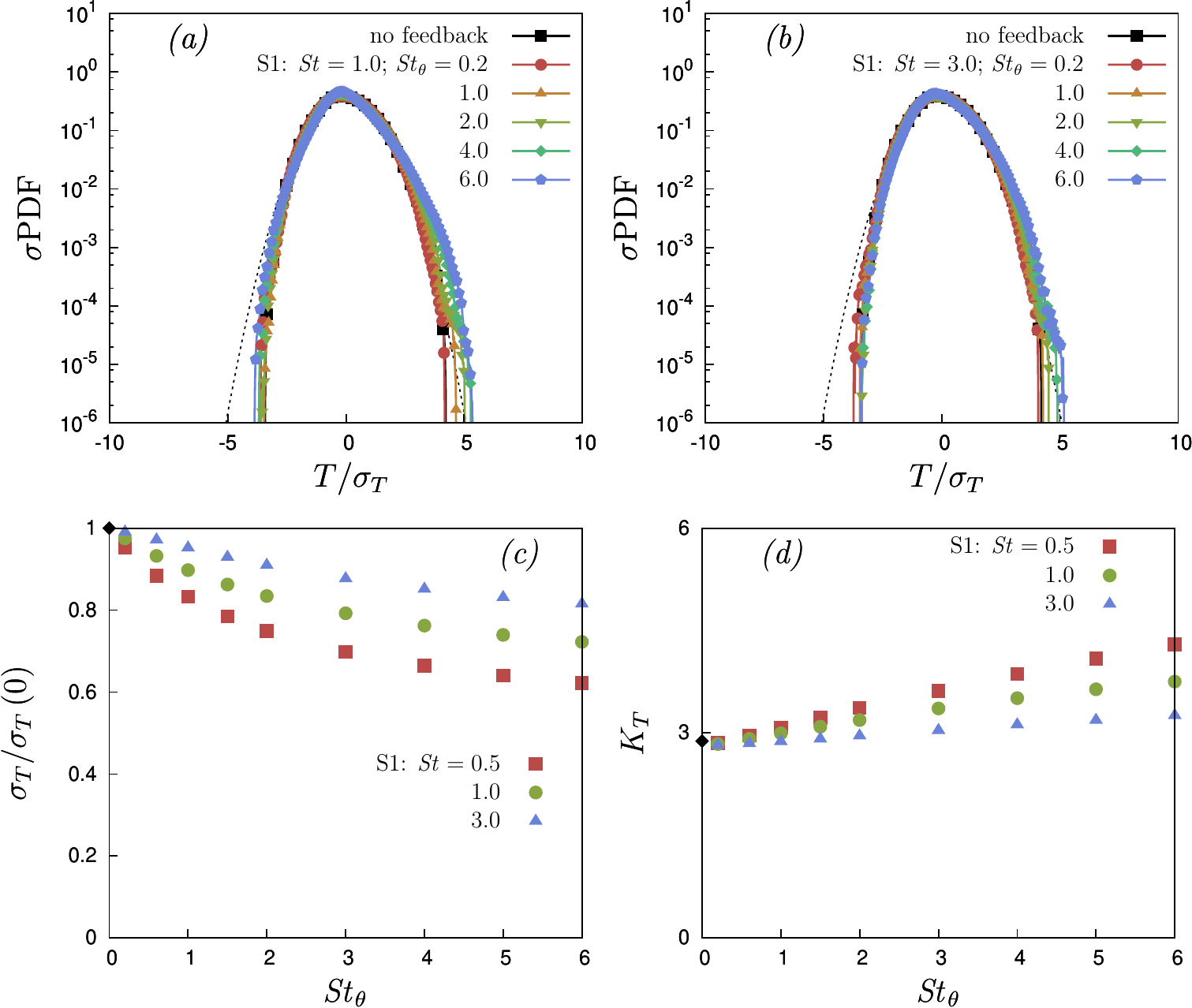}
\caption{PDF of the fluid temperature for $\St=1$ (\textit{a}) and $\St=3$ (\textit{b}), and for various $\St_\theta$. (\textit{c}) Variance of the fluid temperature fluctuations for different $\St$ as a function of $\St_\theta$. (\textit{d}) Kurtosis of the fluid temperature PDF. These results are from simulations S1 in which the two-way thermal coupling is considered.}
\label{fig:PDFT}
\end{figure}

Figures \ref{fig:PDFT}(a-b) show the normalized one-point PDF of the fluid temperature for $\St=1$ and $\St=3$, respectively, and for various $\St_\theta$. The PDFs are normalized with the standard deviation of the distribution $\sigma_{T}$. The PDFs are almost Gaussian for low $\St_\theta$, while the tails become wider as $\St_\theta$ is increased. However, we are unable to explain the cause of this enhanced non-Gaussianity. The temperature PDFs are also not symmetric, and display a bump in the right tail. This behavior was also reported by \citep{Overholt1996} for the case without particles, and it appears to be a low Reynolds number effect that is also dependent on the forcing method employed.

The effect of $\St$ on $\sigma_T$ is striking, whereas we saw earlier in figure \ref{fig:PDFgrT}(c) that $\chi_f$ only weakly depends on $\St$. To explain the dependence upon the Stokes number we note that the energy balance \eqref{eq:dissbal} can be rewritten as
\begin{equation}
\chi = \kappa \left[ \avg{ \left\Vert \bnabla T \right\Vert^2 } + \frac{2}{3}\frac{\phi}{\tau_\eta}\frac{\rho_p}{\rho_0}\frac{1}{\St}\avg{\left( T \of{\vect{x}_p,t} - \theta_p\right)^2 } \right]\label{eq:balSt}.
\end{equation}
The  factor $\phi\rho_p/\left(\rho_0\tau_\eta\right)$ is constant in our simulations. Therefore, since our DNS data suggest that $\chi_f$ is a function of $\St_\theta $ only (see figure \ref{fig:PDFgrT}(c)), from \eqref{eq:balSt} and \eqref{eq:parttheta} we obtain
\begin{equation}
\avg{T\of{\vect{x}_p,t}^2} - \avg{\theta_p^2} \propto \St f\of{\St_\theta}.
\end{equation}

The kurtosis of the fluid temperature fluctuation is shown in figure \ref{fig:PDFT}(d), as a function of $\St_\theta$ and for various $\St$. For small $\St_\theta$, the kurtosis of the fluid temperature fluctuation is close to the value for a Gaussian PDF, namely $3$. However, as $\St_\theta$ is increased, the kurtosis increases. Furthermore, the kurtosis decreases with increasing $\St$ for the range considered in our simulations. The explanation of these trends in the kurtosis is unclear.

\subsection{Particle temperature fluctuations}

\begin{figure}
\centering
\includegraphics[width=\textwidth]{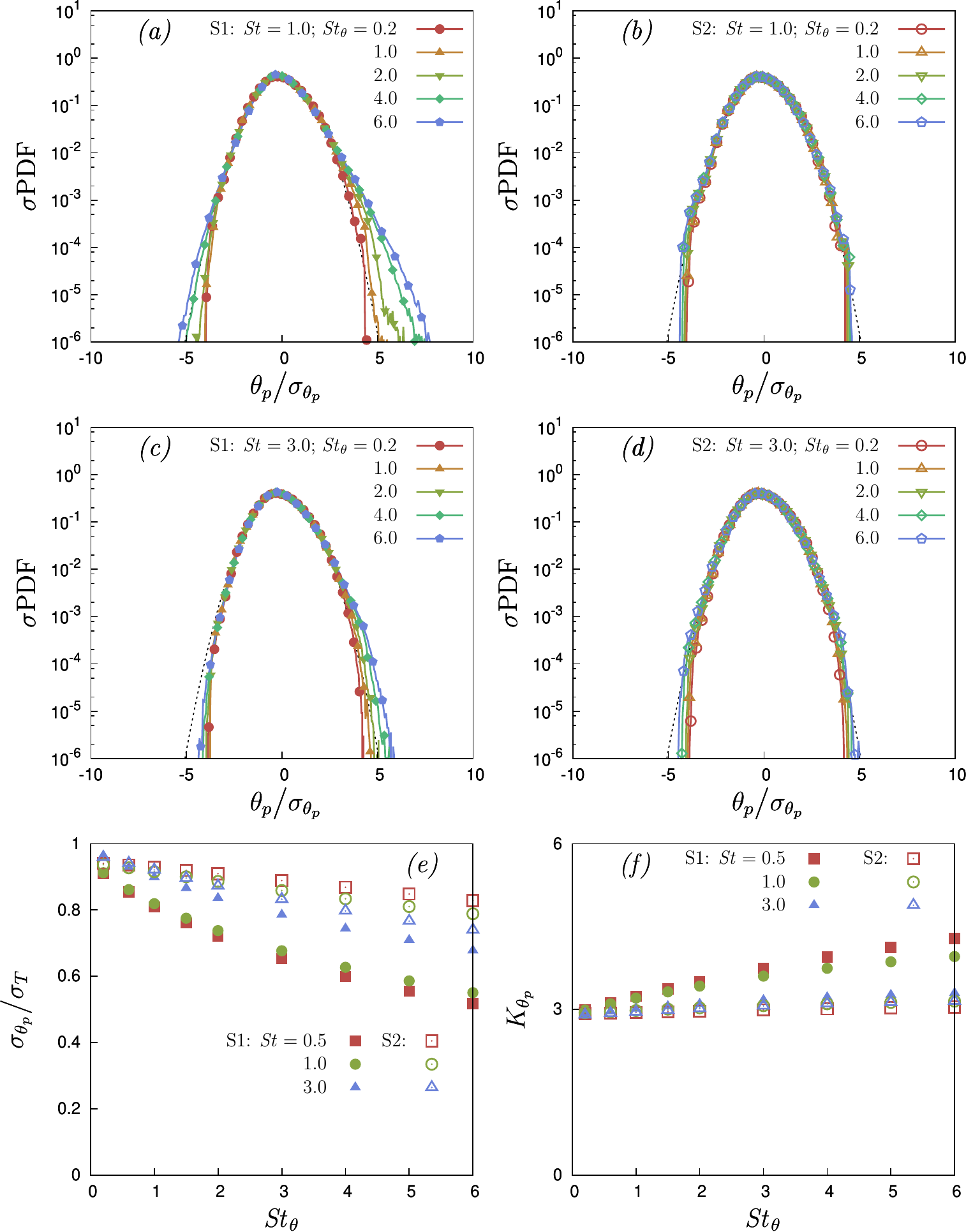}
\caption{PDF of the particle temperature for $\St=1$ (\textit{a-b}) and $\St=3$ (\textit{c-d}), for various $\St_\theta$. Plots (\textit{a-c}) are from simulations S1, in which the two-way thermal coupling is considered, while plots (\textit{b-d}) are from simulations S2, in which the two-way coupling is neglected. (\textit{e}) Variance of the particle temperature fluctuations for different $\St$ numbers as a function of $\St_\theta$. (\textit{f}) Kurtosis of the particle temperature distribution.}
\label{fig:PDFtheta}
\end{figure}

Figures \ref{fig:PDFtheta}(a-b) show the normalized one-point PDF of the particle temperature with $\St=1$, for various $\St_\theta$, and for simulations S1 and S2. Figures \ref{fig:PDFtheta}(c-d) show the corresponding results for $\St=3$, and the PDFs are normalized by their standard deviations. When the two-way thermal coupling is accounted for, the tails of the particle temperature distribution tend to become wider as $\St_\theta$ is increased. On the other hand, when the two-way coupling is neglected, the PDF of the particle temperature is very close to Gaussian, and its shape is not sensitive to either $\St$ or $\St_\theta$.

The variance of the particle temperature fluctuations monotonically decrease with increasing $\St_\theta$, as shown in figure \ref{fig:PDFtheta}(e). The results also show a strong dependence on $\St$, but most interestingly, the dependence on $\St$ is the opposite for the cases with and without two-way coupling. To understand this we note that using the formal solution to the equation for $\dot{\theta}_p(t)$ (ignoring initial conditions) we may construct the result
\begin{equation}
\Big\langle \theta^2_p(t)\Big\rangle=\frac{1}{\tau_\theta^2}\int^t_0\int^t_0 \Big\langle T(\vect{x}_p(s),s)T(\vect{x}_p(s'),s')\Big\rangle  e^{-(2t-s-s')/\tau_\theta}\,ds\,ds'.
\end{equation}
If we now substitute into this the exponential approximation \[\langle T(\vect{x}_p(s),s)T(\vect{x}_p(s'),s')\rangle \approx \langle T^2(\vect{x}_p(t),t)\rangle \exp[-|s-s'|/\tau_T],\] where $\tau_T$ is the timescale of $T(\vect{x}_p(t),t)$, then we obtain
\begin{equation}
\Big\langle \theta^2_p(t)\Big\rangle=   \frac{\langle T^2(\vect{x}_p(t),t)\rangle}{1+\tau_\theta/\tau_T} .
\end{equation}
This result reveals that the particle temperature variance is influenced by $\St$ in two ways. First, $\langle T^2(\vect{x}_p(t),t)\rangle$ depends upon the spatial clustering of the inertial particles, and this depends essentially upon $\St$. Second, the timescale $\tau_T$ is the timescale of the fluid temperature field measured along the inertial particle trajectories, and hence depends upon $\St$. For isotropic turbulence, this timescale is expected to decrease as $\St$ is increased, which would lead to $\langle \theta^2_p(t)\rangle$ decreasing as $\St$ increases, which is the behavior observed in figure \ref{fig:PDFtheta}(e). In the presence of two-way coupling, however, $\langle T^2(\vect{x},t)\rangle$ increases with increasing $\St$, as shown earlier. In the two-way coupled regime this increase in $\langle T^2(\vect{x},t)\rangle$ leads to an increase in $\langle T^2(\vect{x}_p(t),t)\rangle$ that dominates over the decrease of $\tau_T$ with increasing $\St$, and as a result $\langle \theta^2_p(t)\rangle$ increases with increasing $\St$.

The kurtosis of the particle temperature increases with increasing $\St_\theta$ when the two-way thermal coupling is accounted for, as shown in figure \ref{fig:PDFtheta}(f) (simulations S1, filled symbols). Conversely, the kurtosis of the particle temperature remains constant as $\St_\theta$ is increased when the two-way thermal coupling is ignored (simulations S2, open symbols).

\section{Statistics conditioned on the local fluid temperature gradients}
\label{sec:FATT}


In this section we consider additional quantities to obtain deeper insight into the one-point particle to fluid heat flux. In particular, we explore the relationship between this heat flux and the local fluid temperature gradients.

\subsection{Particle clustering on the temperature fronts}

\begin{figure}
\centering
\includegraphics[width=\textwidth]{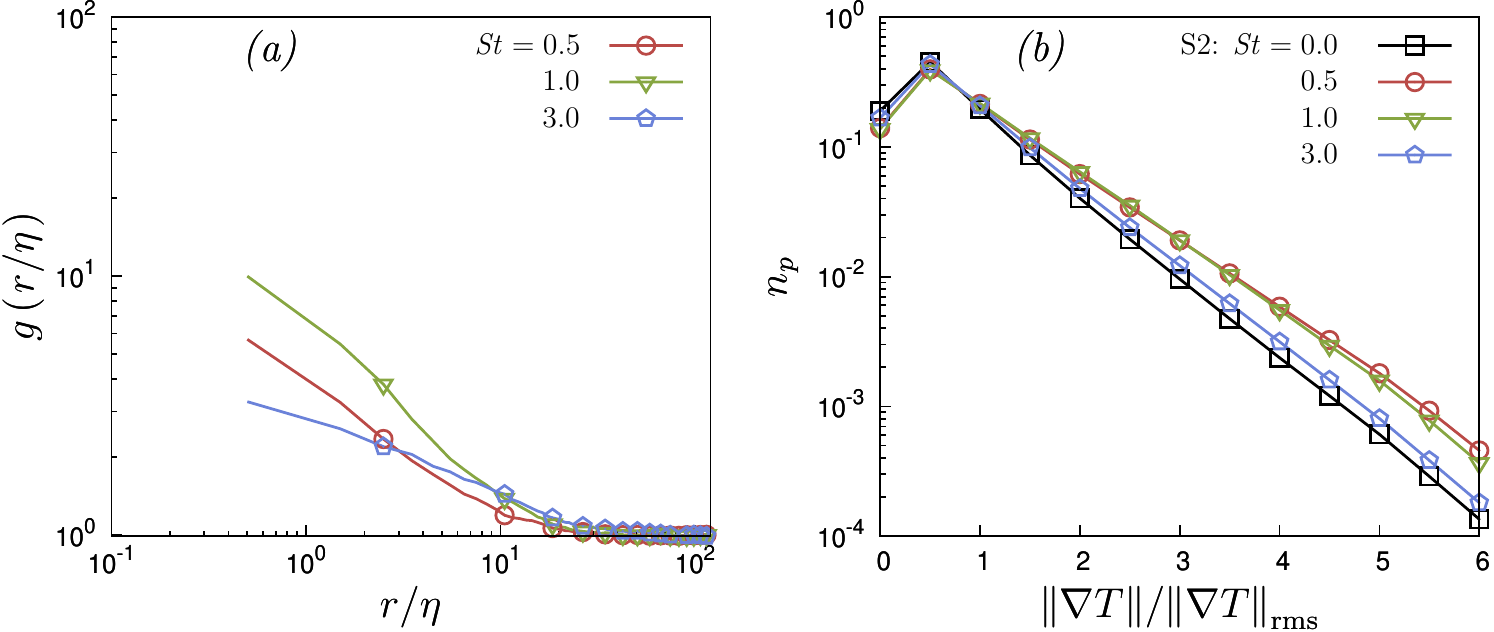}
\caption{(\textit{a}) Radial distribution function (RDF) as a function of the separation $r/\eta$ for various $\St$. (\textit{b}) Particle number density conditioned on the magnitude of the fluid temperature gradient at the particle position, for various $\St$. These results are from simulations S2, in which the two-way thermal coupling is neglected.}
\label{fig:FATTconc}
\end{figure}

It is well known that inertial particles in turbulence form clusters \citep{Bec2007}, which may be quantified using the radial distribution function (RDF). As shown in figure \ref{fig:FATTconc}(a), the particle number density in our simulations at small separations is a order of magnitude larger than the mean density when $\St=\textit{O}\of{1}$. \cite{Bec2014} showed that inertial particles also exhibit a tendency to preferentially cluster in the fluid temperature fronts where the temperature gradients are large. To demonstrate this, they measured the temperature dissipation rate at the particle positions and showed that this was higher than the Eulerian dissipation rate of the fluid temperature fluctuations. Alternatively, we may quantify this tendency for inertial particles to cluster in the fluid temperature fronts by computing the particle number density conditioned on the magnitude of the fluid temperature gradient
\begin{equation}
n_P\of{\left\Vert\bnabla T\right\Vert} = \frac{\sum_p\int_V \delta\of{\vect{x}-\vect{x}_p}\de \vect{x}}{N_P},\; V=\left\{ \vect{x}: \left\Vert\bnabla T\of{\vect{x}}\right\Vert = \left\Vert\bnabla T\right\Vert\right\}.
\end{equation}
Defining $\|\bnabla T\|_{rms}$ as the rms value of $\|\bnabla T\|$, small values of $\|\bnabla T\|/\|\vect \bnabla T\|_{rms}$ may be interpreted as corresponding to the large scales, and are associated with the Lagrangian coherent structures in which the temperature field is almost constant. Large values of $\|\bnabla T\|/\|\bnabla T\|_{rms}$ may be interpreted as corresponding to the small scales, and are associated with fronts in the fluid temperature field.

The results for $n_P$ are shown in figure~\ref{fig:FATTconc}(b), corresponding to simulations without two-way thermal coupling (the results show only a weak dependence on $\St_\theta$ when the two-way coupling is included). For fluid particles, $n_P$ decays almost exponentially with increasing $\|\bnabla T\|$. For values of $\St$ at which the maximum particle clustering takes place, $n_P$ is an order of magnitude larger than the value for fluid particles in regions of strong temperature gradients. These results therefore support the conclusions of \cite{Bec2014} that inertial particles preferentially cluster in the fronts of the fluid temperature field where $\|\bnabla T\|/\|\bnabla T\|_{rms}$ is large.

\subsection{Particle motion across the temperature fronts}
To obtain further insight into the thermal coupling between the particles and fluid we consider the properties of the particle heat flux conditioned on $\|\bnabla T\|$. In particular, we consider the following quantity
\begin{equation}
q_n \of{\left\Vert \bnabla T \right\Vert} \equiv {\of{T\of{\vect{x}_p}-\theta_p}^n\vect{v}_p\bcdot \vect{n}_T\of{\vect{x}_p}}\Big\vert_{\left\Vert \bnabla T \right\Vert},
\end{equation}
where $\mathbf{n}_T$ is the normalized temperature gradient
\begin{equation}
\mathbf{n}_T\of{\vect{x}_p} \equiv \frac{\bnabla T\of{\vect{x}_p}}{\left\Vert \bnabla T\of{\vect{x}_p} \right\Vert}.
\end{equation}

\begin{figure}
\centering
\includegraphics[width=\textwidth]{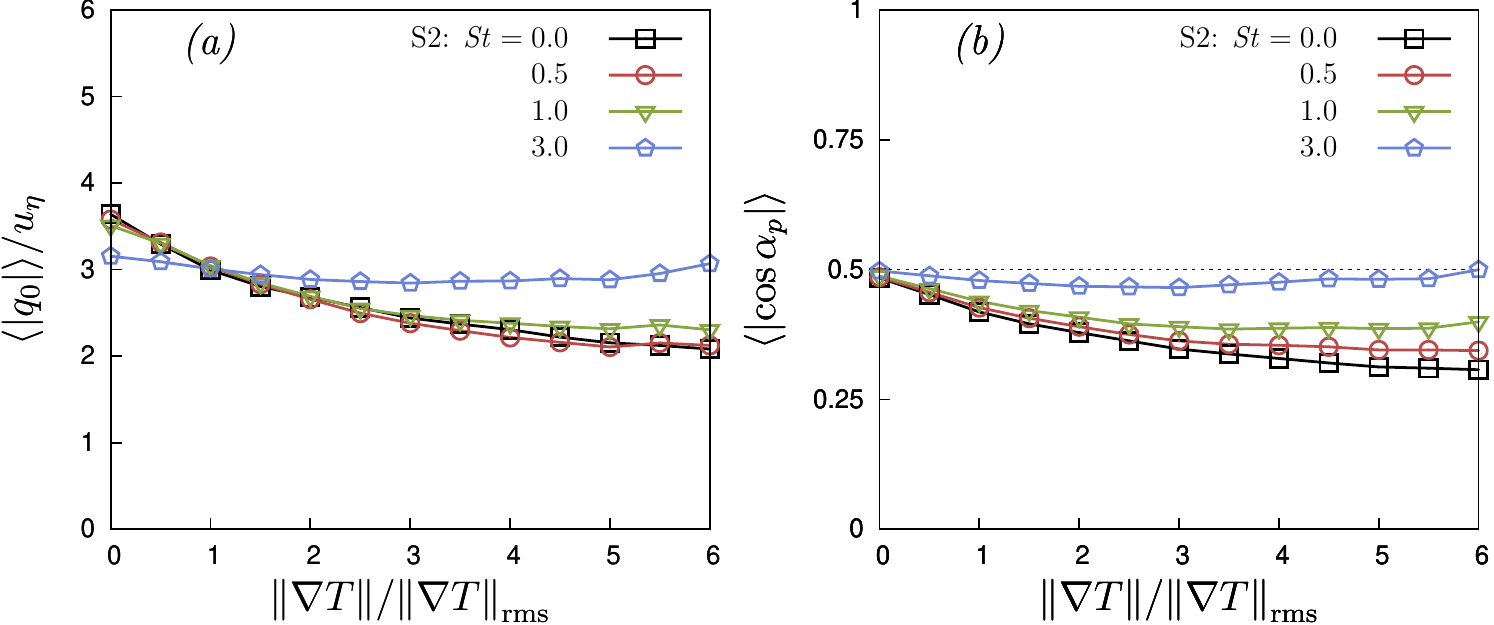}
\caption{(\textit{a}) Results for $\avg{\left\vert q_0(\|\bnabla T\|)\right\vert}/u_\eta$, for various $\St$. (\textit{b}) Results for $\langle|\cos \alpha_p|\rangle$ as a function of $\|\bnabla T\|$, for various $\St$. These results are from simulations S2, in which the two-way thermal coupling is neglected.}
\label{fig:FATTq0}
\end{figure}

The statistics of $q_n$ provide a way to quantify the relationship between the particle heat flux and the local temperature gradients in the fluid. Understanding this relationship is key to understanding how the particles modify the properties of the fluid temperature and temperature gradient fields. 

The efficiency with which the particles cross the fronts in the fluid temperature field is quantified by $\avg{\left\vert q_0\right\vert}$, and our results for this quantity are shown in figure \ref{fig:FATTq0}(a). The curves are normalized with the Kolmogorov velocity scale $u_\eta$. The results show that as $\St$ is increased, the particles move across the fronts with increasingly large velocities. This behavior is non-trivial since it is known that the kinetic energy of an inertial particle decreases with increasing $\St$ \citep{Zaichik2009,Ireland2016}. 

It is also important to consider whether the reduction of $\avg{\left\vert q_0\right\vert}$ as $\|\bnabla T\|$ increases is due to the reduction of the norm of the particle velocity or to the lack of alignment between the particle velocity and the fluid temperature gradient at the particle position. Figure \ref{fig:FATTq0}(b) displays the average of the absolute value of the cosine of the angle between the particle velocity and temperature gradient 
\begin{equation}
\cos\alpha_p \equiv \frac{\vect{v}_p}{\left\Vert \vect{v}_p \right\Vert} \bcdot \frac{\bnabla T \of{\vect{x}_p}}{\left\Vert \bnabla T \of{\vect{x}_p}\right\Vert},
\end{equation}
conditioned on $\|\vect{\bnabla}T\|$.

The results show that as $\|\bnabla T\|$ is increased, the particle motion becomes misaligned with the local fluid temperature gradient. This then shows that the reduction of $\avg{\left\vert q_0\right\vert}$ as $\|\bnabla T\|$ increases is due to non-trivial statistical geometry in the system. The results also show that as $\St$ is increased, the cosine of the angle between the fluid temperature gradient and the particle velocity becomes almost independent of $\|\bnabla T\|$, and $\avg{\left\vert\cos\alpha_p\right\vert}\approx 1/2$, the value corresponding to $\cos\of{\alpha_p}$ being a uniform random variable. This shows that as $\St$ is increased, the correlation between the direction of the particle velocity and the local fluid velocity gradient vanishes.

\subsection{Heat flux due to the particle motion across the fronts}

\begin{figure}
\centering
\includegraphics[width=\textwidth]{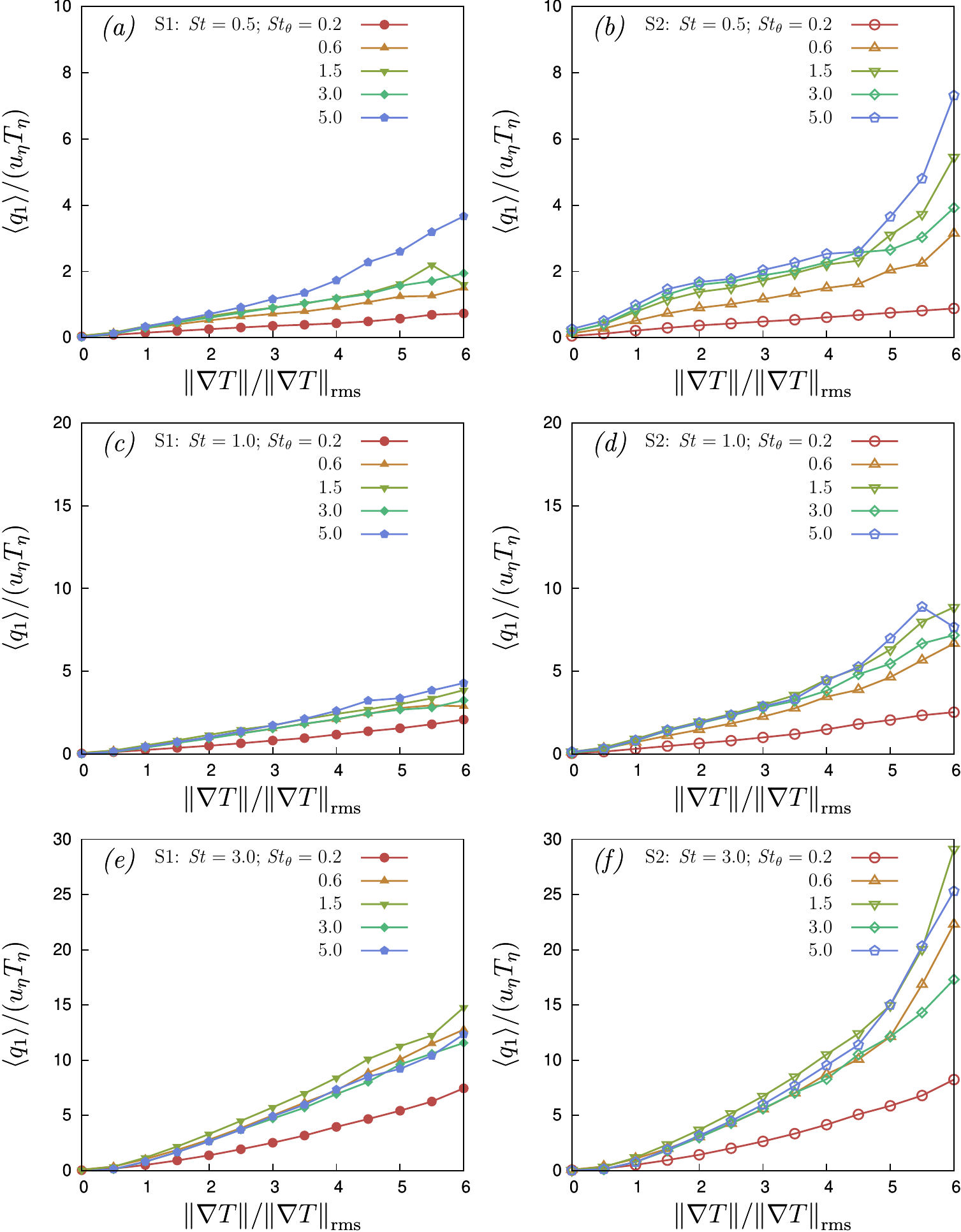}
\caption{Results for  $\avg{q_1\of{\norm{\bnabla T}}}/\of{u_\eta T_\eta}$ for $\St=0.5$ (\textit{a-b}), $\St=1$ (\textit{c-d}) and $\St=3$ (\textit{e-f}), and for various $\St_\theta$. Plots (\textit{a-c-e}) are from simulations S1, in which the two-way thermal coupling is considered, while plots (\textit{b-d-f}) are from simulations S2, in which the two-way coupling is neglected.}
\label{fig:FATTq1}
\end{figure}

We now turn to consider the quantity $\avg{q_1}$. When the particle moves from a cold to a warm region of the fluid, the component of the particle velocity along the temperature gradient is positive, $\vect{v}_p\bcdot \vect{n}_T\of{\vect{x}_p} > 0$. If the particle is also cooler than the local fluid so that $T\of{\vect{x}_p}-\theta_p>0$, then as it moves into the region where the fluid is warmer, $q_1>0$ meaning that the particle will absorb heat from the fluid, and will therefore tend to reduce the local fluid temperature gradient. When the particle moves from a warm to a cold region of the flow, if $T\of{\vect{x}_p}-\theta_p<0$ then $q_1$ is also positive, so that again the particle will act to reduce the local temperature gradient in the fluid. Therefore, $q_1>0$ indicates that the action of the inertial particles is to smooth out the fluid temperature field, reducing the magnitude of its temperature gradients, and $q_1<0$ implies the particles enhance the temperature gradients.

The results for $\avg{q_1}$ are shown in figure \ref{fig:FATTq1} for various $\St$ and $\St_\theta$, including (simulations S1) and neglecting (simulations S2) the two-way thermal coupling. On average we observe $\avg{q_1}\ge 0$, such that the particles tend to make the fluid temperature field more uniform. The results show that $\avg{q_1}$ tends to zero as  $\|\bnabla T\|\to0$. This indicates that the particles spend enough time in the Lagrangian coherent structures to adjust to the temperature of the fluid. However, $\avg{q_1}$ increases significantly as $\|\bnabla T\|$ increases, suggesting that inertial particles can carry large temperature differences across the fronts. In the limit $\St_\theta\to 0$,  $\avg{q_1}\to0$ reflecting the thermal equilibrium between the particles and the fluid. As $\St_\theta$ is increased, the heat-flux becomes finite, however, if $\St_\theta$ is too large, the particle temperature decorrelates from the fluid temperature and the heat exchange is not effective. Hence, $\avg{q_1}$ can saturate with increasing $\St_\theta$. The results show that $\avg{q_1}$ increases with increasing $\St$, associated with the decoupling of $\vect{v}_p$ and $\vect{n}_T\of{\vect{x}_p}$ discussed earlier. Finally, the results also show that two-way thermal coupling reduces $\avg{q_1}$. This is simply a reflection of the fact that since the particles tend to smooth out the fluid temperature gradients, the disequilibrium between the particle and local fluid temperature is reduced, which in turn reduces the heat flux due to the particles.

\section{Temperature structure functions}
\label{sec:SF}

We now turn to consider two-point quantities in order to understand how the two-way thermal coupling affects the system at the small scales.

\subsection{Fluid temperature structure functions}

\begin{figure}
\centering
\includegraphics[width=\textwidth]{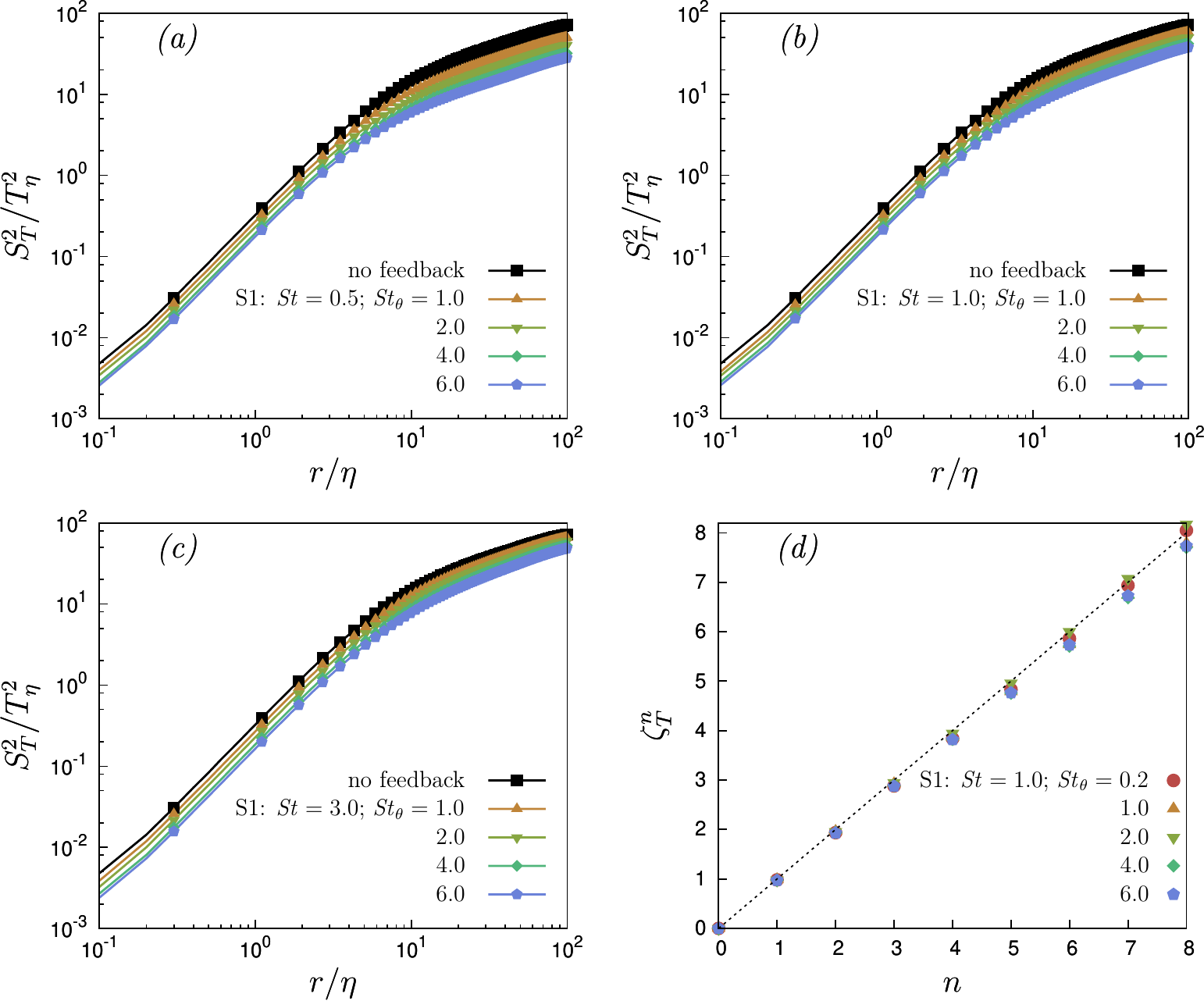}
\caption{Results for $S^2_T$ for different $\St_\theta$, for $\St=0.5$ (\textit{a}), $\St=1$ (\textit{b}) and $\St=3$ (\textit{c}). (\textit{d}) Scaling exponents of the fluid temperature structure functions at small separation, $r\le2\eta$, at $\St=1$. The data is from simulations S1 in which the two-way thermal coupling is considered.}
\label{fig:SFT}
\end{figure}

The $n$-th order structure function of the fluid temperature field is defined as
\begin{equation}
S^n_T\of{r} \equiv \avg{\left\vert \Delta T(r,t)\right\vert^n}
\end{equation}
where $\Delta T(r,t)$ it the difference in the temperature field at two points separated by the distance $r$ (the ``temperature increment''). The results for $S^2_T$, with different $\St$ and $\St_\theta$ are shown in figure \ref{fig:SFT}.

The results show that $S^2_T$ decreases monotonically with increasing $\St_\theta$ at all scales when the two-way thermal coupling is taken to account. In the dissipation range, $S^2_T$ is directly connected to the dissipation rate, and is suppressed in the same way for the three different $\St$ considered. Conversely, the suppression of the large scale fluctuations is stronger as $\St$ is reduced, at least for the range of $\St$ considered here.

The scaling exponents of the structure functions of the temperature field
\begin{equation}
\zeta^n_T \equiv \frac{\de \log S^n_T \of{r}}{\de \log r}
\end{equation}
are shown in figure \ref{fig:SFT}(d) for $r\le2\eta$. The results show that the fluid temperature field remains smooth (to within numerical uncertainty) even when suspended particles are suspended in the flow. This is not trivial since the contribution from the particle to fluid coupling term $C_T \left(\vect{x}+\vect{r},t\right)-C_T \left(\vect{x},t\right)$ need not be a smooth function of $r$.

\subsection{Particle temperature structure functions}

\begin{figure}
\centering
\includegraphics[width=\textwidth]{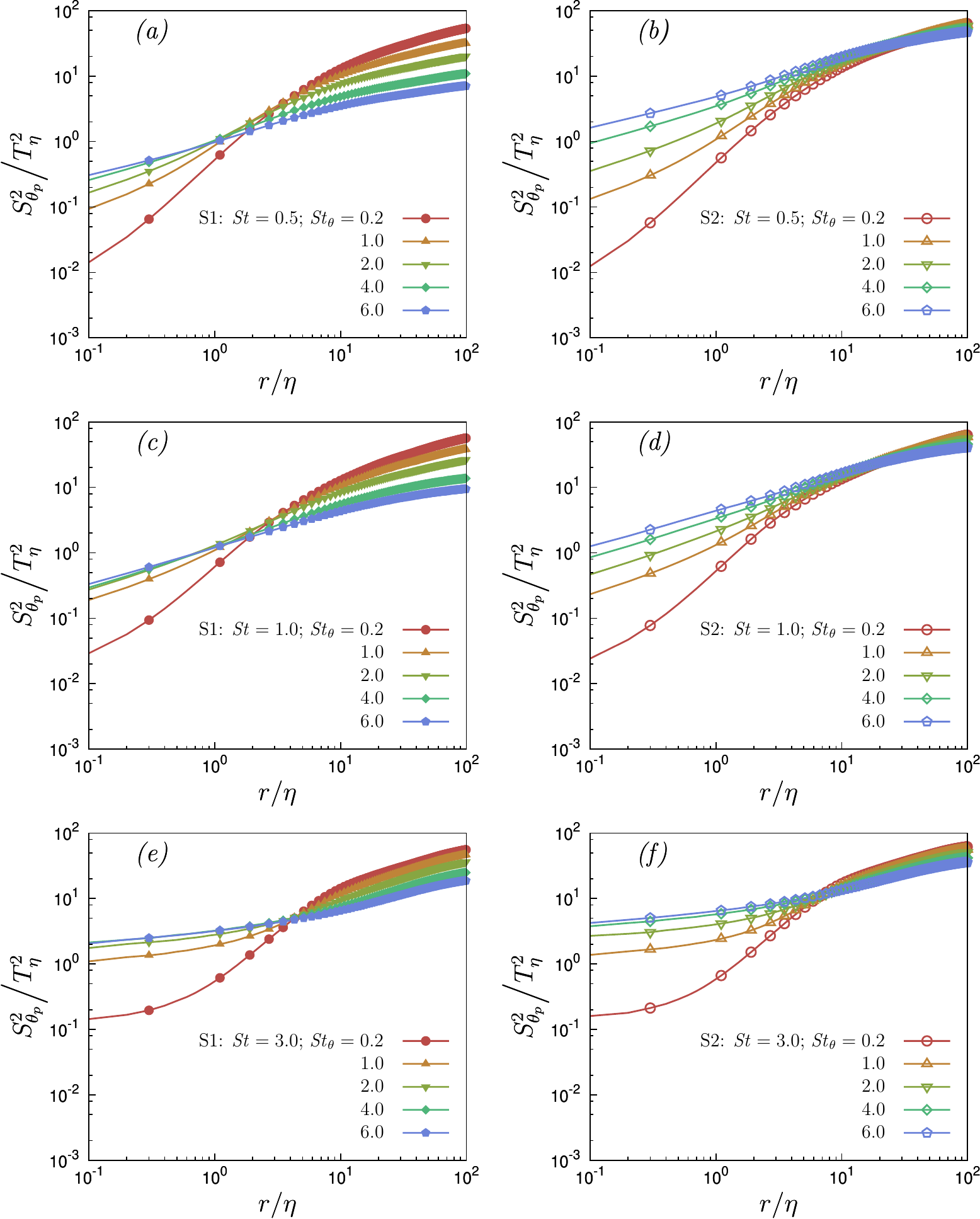}
\caption{Results for $S^2_\theta$ for different $\St_\theta$, for $\St=0.5$ (\textit{a-b}), $\St=1$ (\textit{c-d}) and $\St=3$ (\textit{e-f}). Plots (\textit{a-c-e}) are from simulations S1, in which the two-way thermal coupling is considered, while plots (\textit{b-d-f}) are from simulations S2, in which the two-way coupling is neglected.}
\label{fig:SFtheta}
\end{figure}

The $n$-th order structure function of the particle temperature $\theta_p\of{t}$ is defined as
\begin{equation}
S^n_{\theta}\of{r} \equiv \avg{\left\vert \Delta \theta_{p}\right\vert^n}_{ r}
\end{equation}
where $\Delta \theta_{p}(t)$ is the difference in the temperature of the two particles, and the brackets denote an ensemble average, conditioned on the two particles having separation $r$. The results for $S^2_{\theta}$ for different $\St$ and $\St_\theta$, with and without two-way thermal coupling, are shown in figure \ref{fig:SFtheta}.

The results show that $S^2_{\theta}$ depends on $\St_\theta$ in much the same way as the inertial particle relative velocity structure functions depend on $\St$ \citep{Ireland2016}. This is not surprising since the equation governing $\dot{\theta}_p$ is structurally identical to the equation governing the particle acceleration. However, important differences are that  $\dot{\theta}_p$ depends on both $\St$ and $\St_\theta$, and also that the fluid temperature field is structurally different from the fluid velocity field, with the temperature field exhibiting the well-known ramp-cliff structure.

To obtain further insight into the behavior of $S^2_{\theta}$  and $S^n_{\theta}$ in general, we note that the formal solution for $\Delta \theta_{p}(t)$ is given by (ignoring initial conditions)
\begin{equation}
\Delta\theta_p\of{t} =\frac{1}{\tau_\theta} \int_{0}^t  \Delta T\of{\vect{x}_p\of{s},\vect{r}_p\of{s},s} \exp\of{-\frac{t-s}{\tau_\theta}} \de s,
\label{eq:soltheta}
\end{equation}
where $\Delta T\of{\vect{x}_p\of{s},\vect{r}_p\of{s},s}$ is the difference in the fluid temperature at the two particle positions $\vect{x}_p\of{s}$ and $\vect{x}_p\of{s}+\vect{r}_p\of{s}$. Equation \eqref{eq:soltheta} shows that $\Delta\theta_p\of{t}$ depends upon $\Delta T$ along the path-history of the particles, and $\Delta\theta_p\of{t}$ is therefore a non-local quantity. The role of the path-history increases as $\St_\theta$ is increased since the exponential kernel in the convolution integral decays more slowly as $\tau_\theta$ is increased. Since the statistics of $\Delta T$ increase with increasing separation, particle-pairs at small separations are able to be influenced by larger values of $\Delta T$ along their path-history, such that $\Delta\theta_p\of{t}$ can significantly exceed the local fluid temperature increment $\Delta T\of{\vect{x}_p\of{t},\vect{r}_p\of{t},t}$. This then causes $S^2_{\theta}$ to increase with increasing $\St_\theta$, as shown in figure \ref{fig:SFtheta}. This effect is directly analogous to the phenomena of caustics that occur in the relative velocity distributions of inertial particles at the small scales of turbulence \citep{Wilkinson2005}, and which occur because the inertial particle relative velocities depend non-locally on the fluid velocity increments experienced along their trajectory history \citep{bragg14c}. In analogy, we may therefore refer to the effect as ``thermal caustics'', and they may be of particular importance for particle-laden turbulent flows where particles in close proximity thermally interact.

The results in figure \ref{fig:SFtheta} also reveal a strong effect of $\St$, and one way that $\St$ affects these results is through the spatial clustering and preferential sampling of the fluid temperature field by the inertial particles. There is, however, another mechanism through which $\St$ can affect $S^2_{\theta}$. In particular, since, due to caustics, the relative velocity of the particles increases with increasing $\St$ at the small scales, then the values of $\Delta T\of{\vect{x}_p\of{s},\vect{r}_p\of{s},s}$ that may contribute to $\Delta\theta_p\of{t}$ become larger. This follows since if their relative velocities are larger, then over the time span $t-s\leq \orderof{\tau_\eta}$ the particle-pair can come from even larger scales where (statistically) $\Delta T\of{\vect{x}_p\of{s},\vect{r}_p\of{s},s}$ is bigger. This effect would cause $S^2_{\theta}$ to increase with $\St$ for a given $\St_\theta$, further enhancing the thermal caustics, which is exactly what is observed in figure \ref{fig:SFtheta}. The results also show that the thermal caustics are stronger for $\St_\theta\geq \orderof{1}$ when the two-way thermal coupling is ignored. This is mainly due to the reduction in the fluid temperature gradients due to the two-way thermal coupling described earlier, noting that in the limit of vanishing fluid temperature gradients, the thermal caustics necessarily disappear.

At larger scales where the statistics of $\Delta T$ vary more weakly with $r$, the non-local effect weakens, the thermal caustics disappear, and a filtering mechanism takes over which causes $S^2_{\theta}$ to decrease with increasing $\St_\theta$. This filtering effect is directly analogous to that dominating the large-scale velocities of inertial particles in isotropic turbulence, and is associated with the sluggish response of the particles to the large scale flow fluctuations due to their inertia \citep{Ireland2016}.

\begin{figure}
\centering
\includegraphics[width=\textwidth]{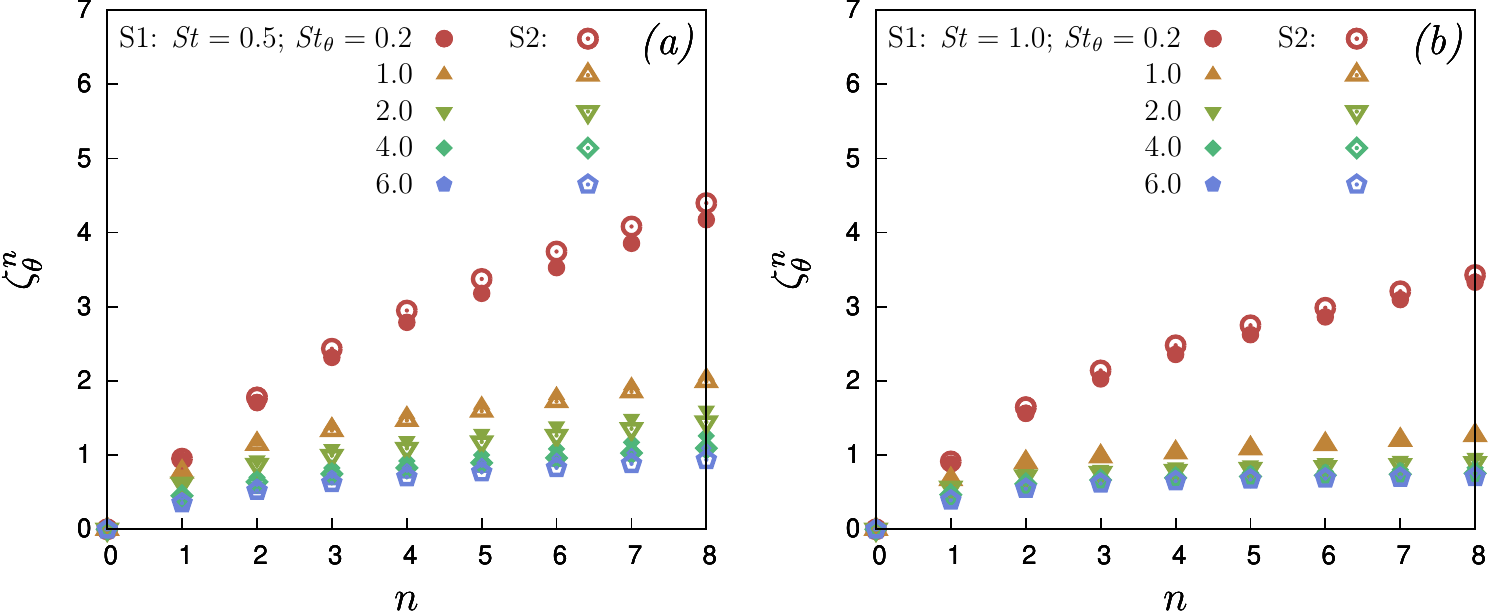}
\caption{Scaling exponent of the structure functions of the particle temperature at small separation, $r\le2\eta$, for various thermal Stokes numbers $\St_\theta$, at $\St=0.5$ (\textit{a}) and $\St=1$ (\textit{b}).}
\label{fig:scalexpSFtheta}
\end{figure}

In the dissipation range our results show that $S^n_{\theta}$ behave as power laws, and the associated scaling exponents $\zeta^n_\theta$ are shown in figure \ref{fig:scalexpSFtheta}. To reduce statistical noise, we estimate $\zeta^n_\theta$ by fitting the data for $S^n_{\theta}$  over the range $r\leq 2\eta$. Over this range, $S^n_{\theta}$ do not strictly behave as power laws, and hence the exponents measured are understood as average exponents. The results in figure \ref{fig:scalexpSFtheta} reveal that particle temperature increments exhibit a strong multifractal behaviour. This multifractility is due to the non-local thermal dynamics of the particles and the formation of thermal caustics, described earlier. In particular, there exists a finite probability to find inertial particle-pairs that are very close but have large temperature differences because they experienced very different fluid temperatures along their trajectory histories. As with the thermal caustics, the multifractility is enhanced as $\St$ is increased. 

Most interestingly, the results for $\zeta^n_\theta$ are only weakly affected by the two-way thermal coupling, despite the fact that we observed a significant effect of the coupling on $S^2_{\theta}$. This suggests that the two-way coupling affects the strength of the thermal caustics, but only weakly affects the scaling of the structure functions in the dissipation range.

\subsection{Mixed structure functions}

\begin{figure}
\centering
\includegraphics[width=\textwidth]{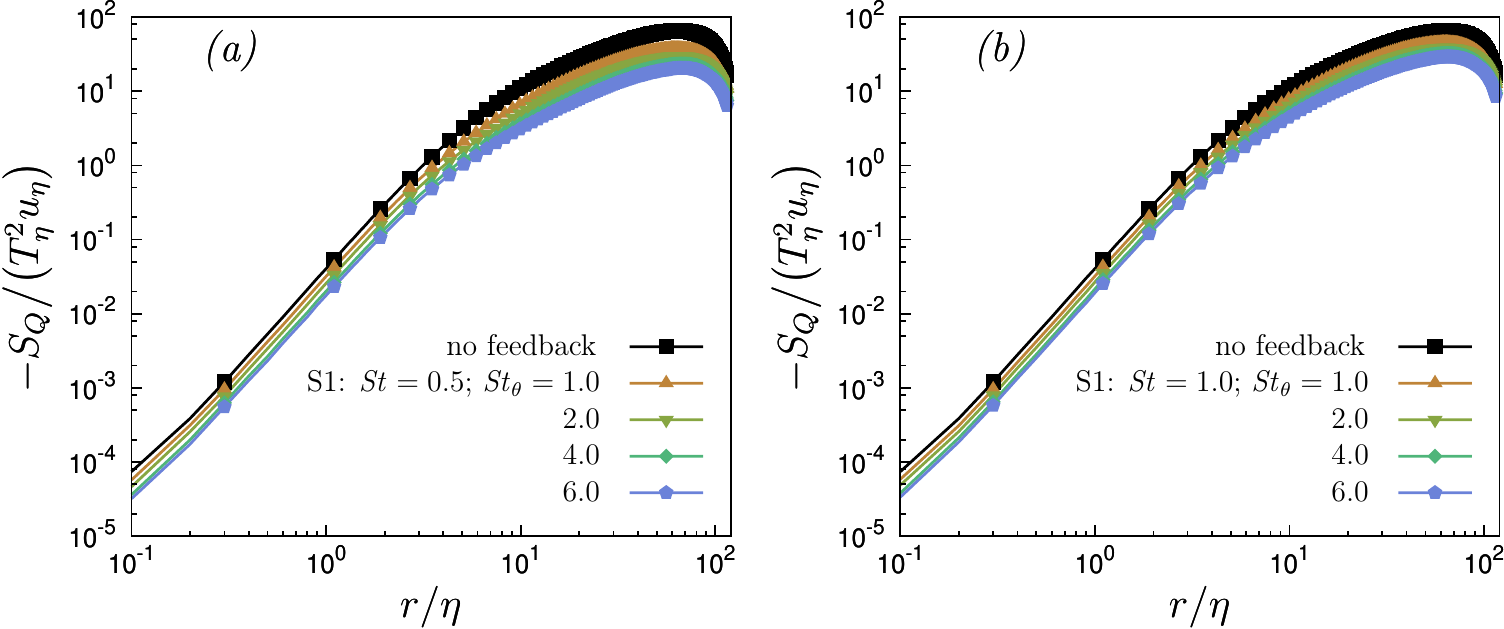}
\caption{Second order mixed structure functions of the fluid temperature field, for different thermal Stokes numbers of the suspended particles, at $\St=0.5$ (\textit{a}) and $\St=1$ (\textit{b}). The data refer to the set of simulations S1, with thermal particle back-reaction included.}
\label{fig:mixedSFT}
\end{figure}

We turn to consider the behaviour of the flux of the temperature increments across the scales of the flow, which is associated with the mixed structure functions
\begin{equation}
S_Q (r)\equiv \avg{\left(\Delta T(r,t)\right)^2 \Delta u_\parallel(r,t)}
\end{equation}
where $\Delta u_\parallel$ is the longitudinal relative velocity difference. The results for $S_Q$, for different $\St$ and $\St_\theta$ are shown in figure \ref{fig:mixedSFT}. Just as we observed for the fluid temperature structure functions, $-S_Q$ decreases monotonically with increasing $\St_\theta$, as was also observed for the fluid temperature dissipation rate $\chi_f$. 

\begin{figure}
\centering
\includegraphics[width=\textwidth]{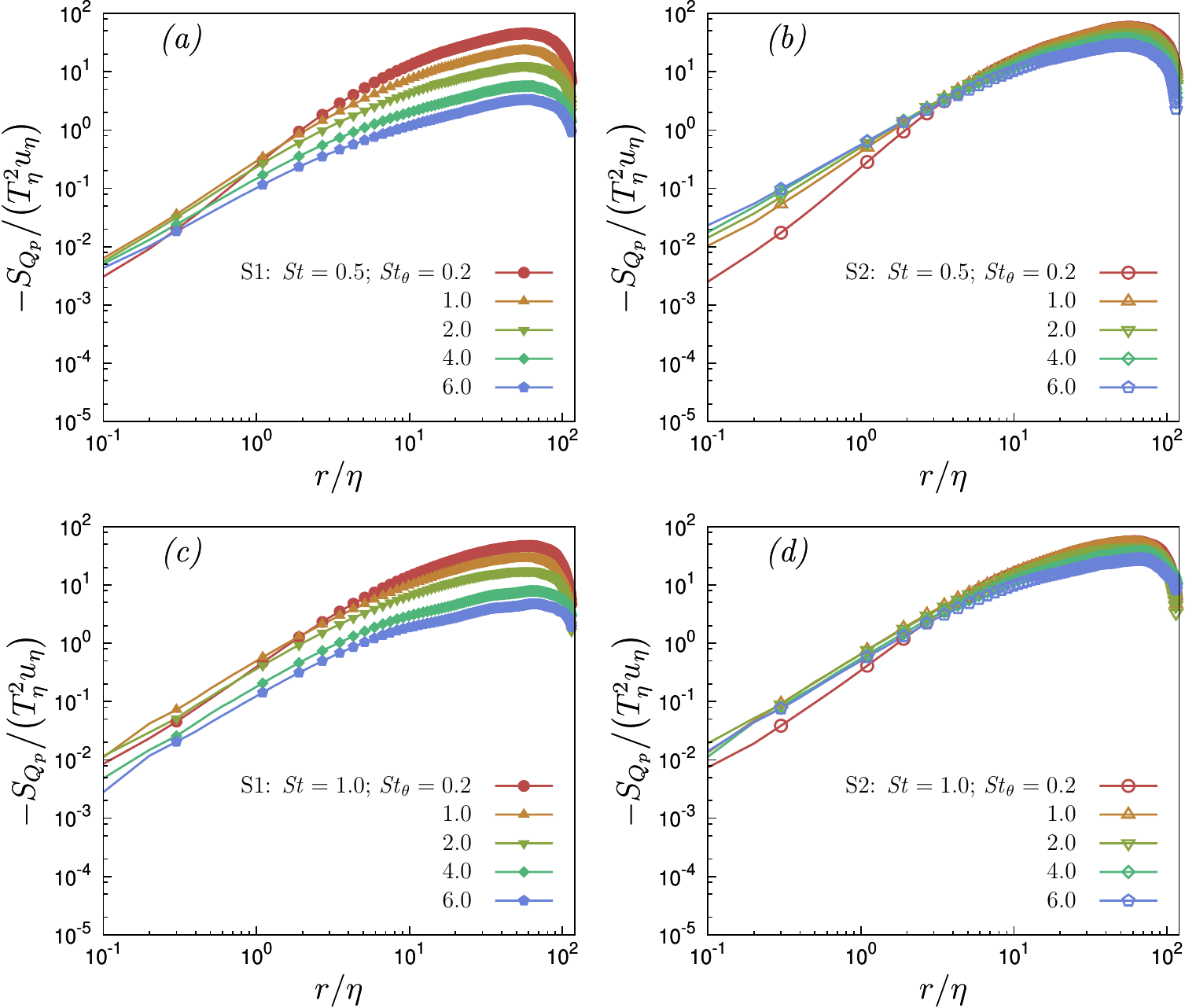}
\caption{Second order mixed structure functions of the particle temperature, for different thermal Stokes numbers, at $\St=0.5$ (\textit{a-b}) and $\St=1$ (\textit{c-d}). The plots on the left (\textit{a-c}) refer to the set of simulations S1, in which the thermal particle back-reaction is included. The plots on the right (\textit{b-d}) refer to the set of simulations S2, in which the thermal particle back-reaction is neglected.}
\label{fig:mixedSFtheta}
\end{figure}

To consider the flux of the particle temperature increments, we begin by considering the exact equation that can be constructed for $S^n_\theta$ using PDF transport equations. In particular, if we introduce the PDF $\mathcal{P}(\vect{r},\Delta\theta,t)\equiv\langle \delta(\vect{r}_p(t)-\vect{r})\delta(\Delta\theta_p(t)-\Delta\theta)\rangle$ and the associated marginal PDF $\varrho(\vect{r},t)\equiv \int\mathcal{P}\,d\Delta\theta$, where $\vect{r}$ and $\Delta\theta$ are time-independent phase-space coordinates, then we may derive for a statistically stationary system  the result (see \cite{bragg14b,bragg2015a} for details on how to derive such results)
\begin{equation}
\Big\langle [\Delta\theta_p(t)]^2\Big\rangle_\vect{r}=\Big\langle \Delta T(\vect{x}_p(t)\vect{r}_p(t),t)\Delta\theta_p(t)\Big\rangle_\vect{r}-\frac{\tau_\theta}{2\varrho}\frac{\partial}{\partial\vect{r}}\bcdot\varrho \Big\langle [\Delta\theta_p(t)]^2\vect{w}_p(t)\Big\rangle_\vect{r},
\end{equation}
where $\vect{w}_p(t)\equiv \partial_t\vect{r}_p(t)$. The first term on the right-hand side is the local contribution that remains when there exist no fluxes across the scales, and this term determines the behavior of $\langle [\Delta\theta_p(t)]^2\rangle_\vect{r}$ at the large scales of homogeneous turbulence where the statistics are independent of $\vect{r}$. The second term on the right-hand side is the non-local contribution that arises for $\St_\theta>0$, and it is this term that is responsible for the thermal caustics discussed earlier. It depends on the spatial clustering of the particles through $\varrho$ (which is proportional to the RDF), and the flux $\langle [\Delta\theta_p(t)]^2\vect{w}_p(t)\rangle_\vect{r}$ which, for an isotropic system, is determined by the longitudinal component
\begin{equation}
S_{Q_p}  (r)\equiv \frac{\vect{r}}{r}\bcdot\Big\langle [\Delta\theta_p(t)]^2\vect{w}_p(t)\Big\rangle_{r}.
\end{equation}
The results for $S_{Q_p}$ from our simulations are shown in figure \ref{fig:mixedSFtheta}, and they show that without two-way coupling, $-S_{Q_p}$ monotonically increases with increasing $\St_\theta$ at the smallest scales. However, with two-way coupling, $-S_{Q_p}$ is maximum for intermediate values of $\St_\theta$, and this occurs because as shown earlier, as $\St_\theta$ is increased, the fluid temperature fluctuations are suppressed across the scales.

\section{Distribution of the temperature increments and fluxes}
\label{sec:2pPDF}

In this section we look at the distribution of the fluid and particle temperature increments in the dissipation range. 
\begin{figure}
\centering
\includegraphics[width=\textwidth]{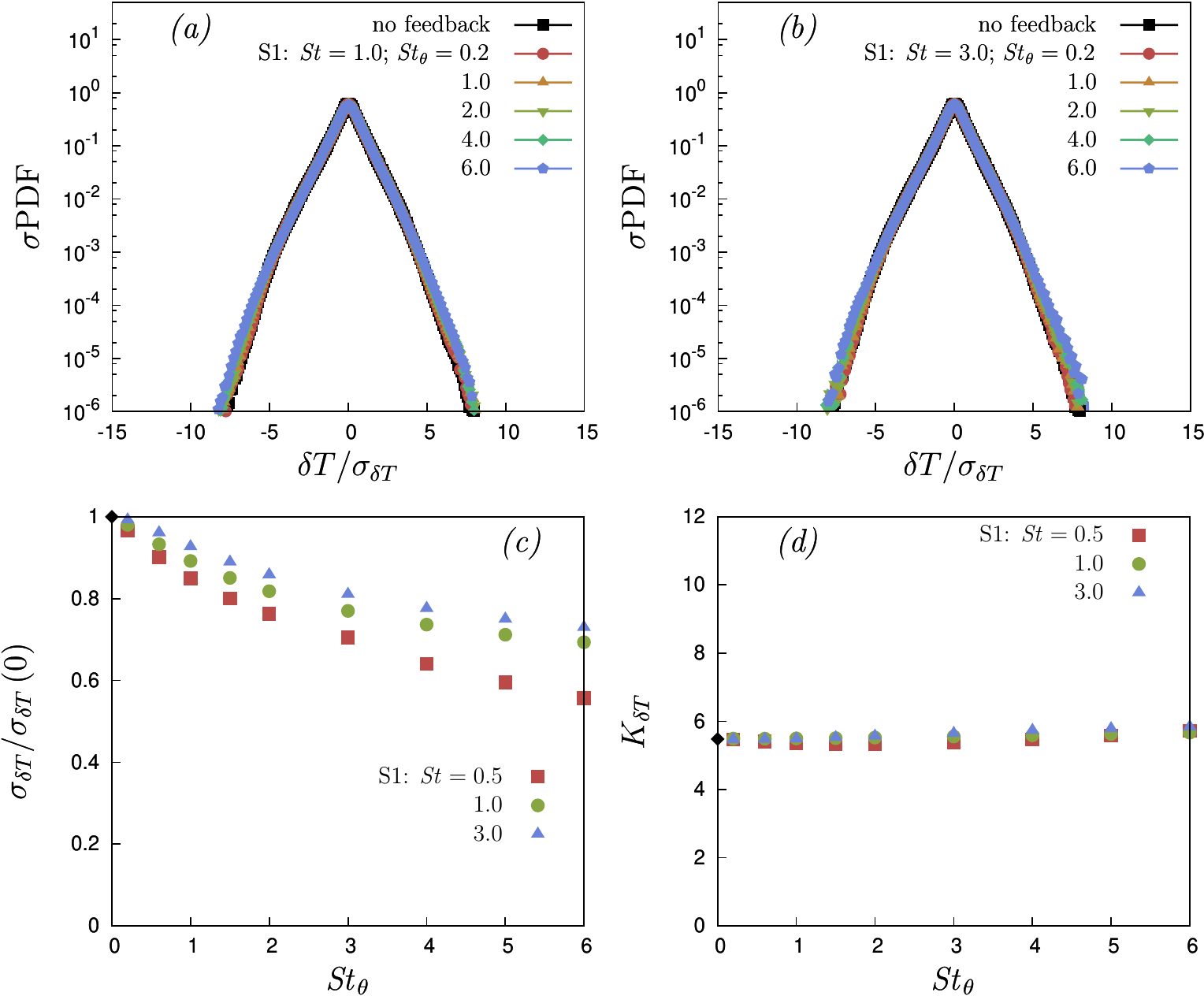}
\caption{Probability density function in normal form of the fluid temperature increments at small separations, $r=10\eta$, at $\St=1$ (\textit{a}) and $\St=3$ (\textit{b}). The data refer to the set of simulations S1, with thermal feedback included. (\textit{c}) Standard deviation of the fluid temperature increments at small separation. (\textit{d}) Kurtosis of the distribution of the fluid temperature increments at small separation.}
\label{fig:PDFdT}
\end{figure}

\subsection{Temperature increments in the dissipation range}

The normalized PDF of the fluid temperature increments at separations $r=10\eta$ are shown in figure \ref{fig:PDFdT}  (for the fluid temperature field, we do not consider the PDFs of the velocity increments for $r\leq \orderof{\eta}$ since these are essentially identical to the PDFs of the fluid temperature gradients that were considered earlier). Just as we observed earlier for the PDFs of the fluid temperature gradients, the results in figure \ref{fig:PDFdT} show that at larger separations the PDFs of the fluid temperature increments are also self similar and approximately collapse when scaled by their standard deviation. The standard deviation and kurtosis of the PDF, also shown in \ref{fig:PDFdT}, show that while the kurtosis is almost independent of $\St$ and $\St_\theta$, the variance decreases with increasing $\St_\theta$, and increases with increasing $\St$. This latter result differs significantly from the behavior of the variance of the fluid temperature gradients which were almost independent of $\St$.
\begin{figure}
\centering
\includegraphics[width=\textwidth]{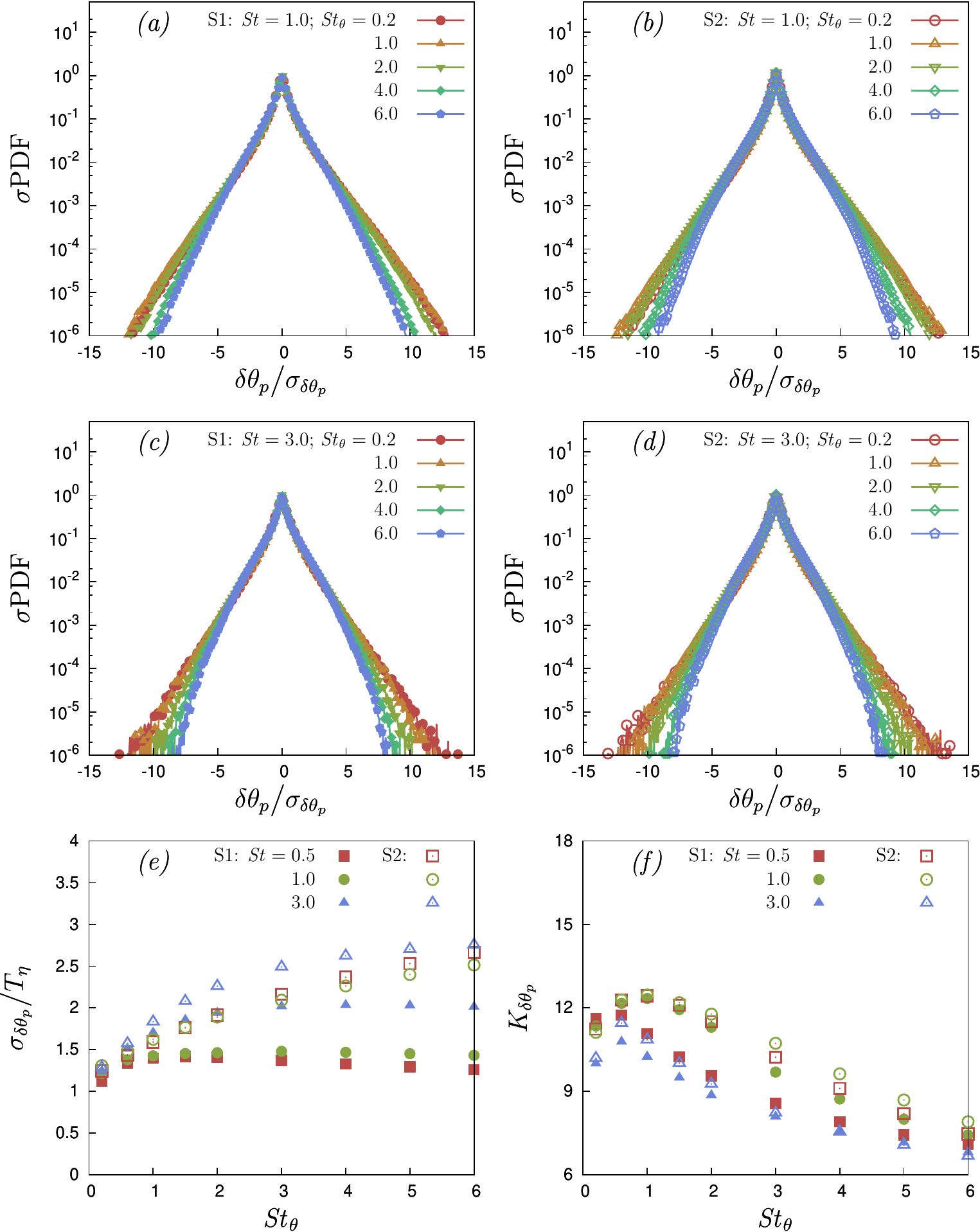}
\caption{Probability density function in normal form of the particle temperature increments at small separations, $r\le2\eta$, at $\St=1$ (\textit{a-b}) and $\St=3$ (\textit{c-d}). The plots on the left (\textit{a-c}) refer to the set of simulations S1, in which the thermal particle back-reaction is included. The plots on the right (\textit{b-d}) refer to the set of simulations S2, in which the thermal particle back-reaction is neglected. (\textit{e}) Standard deviation of the particle temperature increments at small separation. (\textit{f}) Kurtosis of the distribution of the particle temperature increments at small separation.}
\label{fig:PDFdtheta}
\end{figure}

The PDF of the particle temperature increments, along with its variance and kurtosis are shown in figure \ref{fig:PDFdtheta}. The results show that while the variance of the PDF monotonically increases with increasing $\St_\theta$, the kurtosis can increase slightly with increasing $\St_\theta$ when $\St_\theta$ is below some threshold, after which the kurtosis monotonically decreases with increasing $\St_\theta$. However, across the parameter range studied, the PDFs are strongly non-Gaussian, with a maximum kurtosis value of $\approx 13$. The kurtosis values are also strongly and non-monotonically dependent on $\St$, with the largest values tending to occur for $\St=1$. This may be due to the clustering of the particles in the fronts of the temperature field, leading to large particle temperature differences. It may also be due to the non-local mechanisms described earlier since although the non-local effects can enhance non-Gaussianity in certain regimes, in the regime where the behavior is entirely non-local (e.g. for $\St\gg1$), the behavior becomes ballistic and $\Delta\theta_p(t)$ is governed by a central limit theorem and the PDF of $\Delta\theta_p(t)$ approaches a Gaussian distribution.

\subsection{Flux of temperature increments in the dissipation range}

We finally turn to consider the PDFs of the fluid temperature flux $Q=\left(\Delta T(r,t)\right)^2 \Delta u_\parallel(r,t)$ and particle temperature flux $Q=[\Delta\theta_p(t)]^2{w}_\parallel(t)$, where ${w}_\parallel(t)$ is the parallel component of the particle-pair relative velocity.

\begin{figure}
\centering
\includegraphics[width=\textwidth]{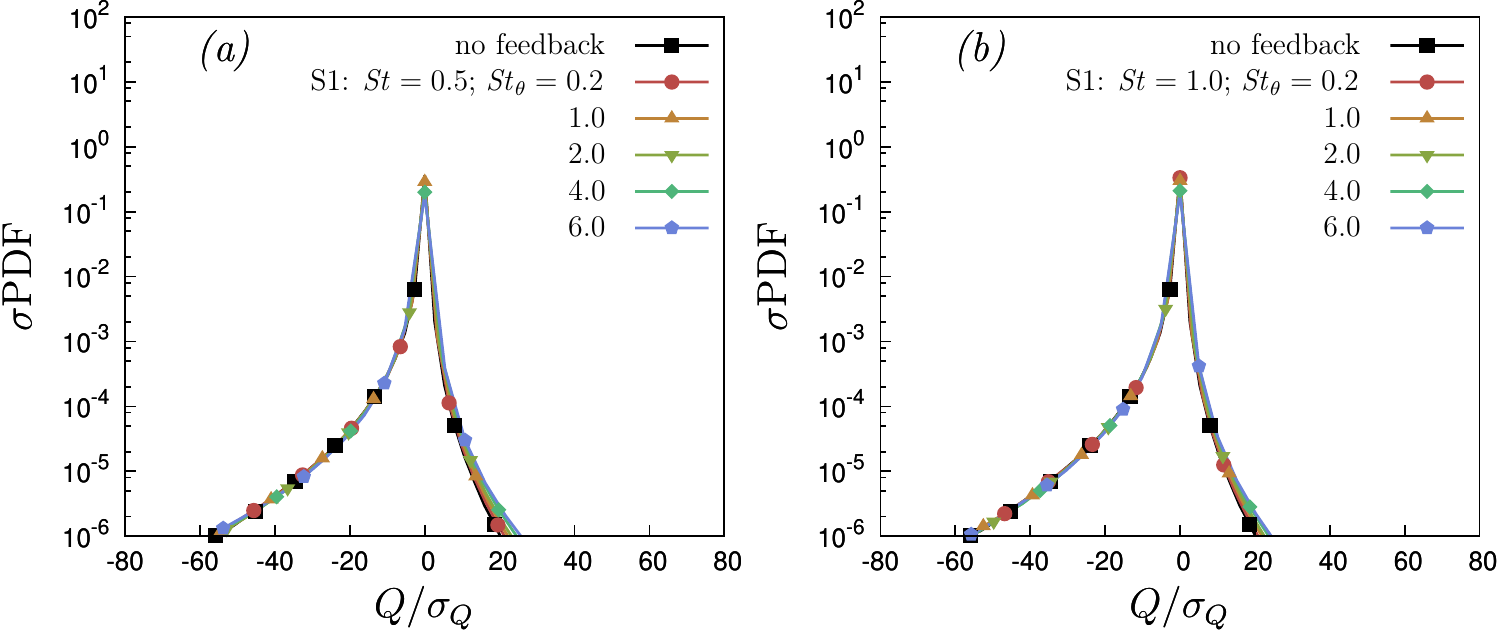}
\caption{Probability density function in normal form of the flux of fluid temperature increments at small separations, $r\le2\eta$, at $\St=0.5$ (\textit{a}) and $\St=1$ (\textit{b}). The data refer to the set of simulations S1, with thermal feedback included.}
\label{fig:PDFfluxdT}
\end{figure}

\begin{figure}
\centering
\includegraphics[width=\textwidth]{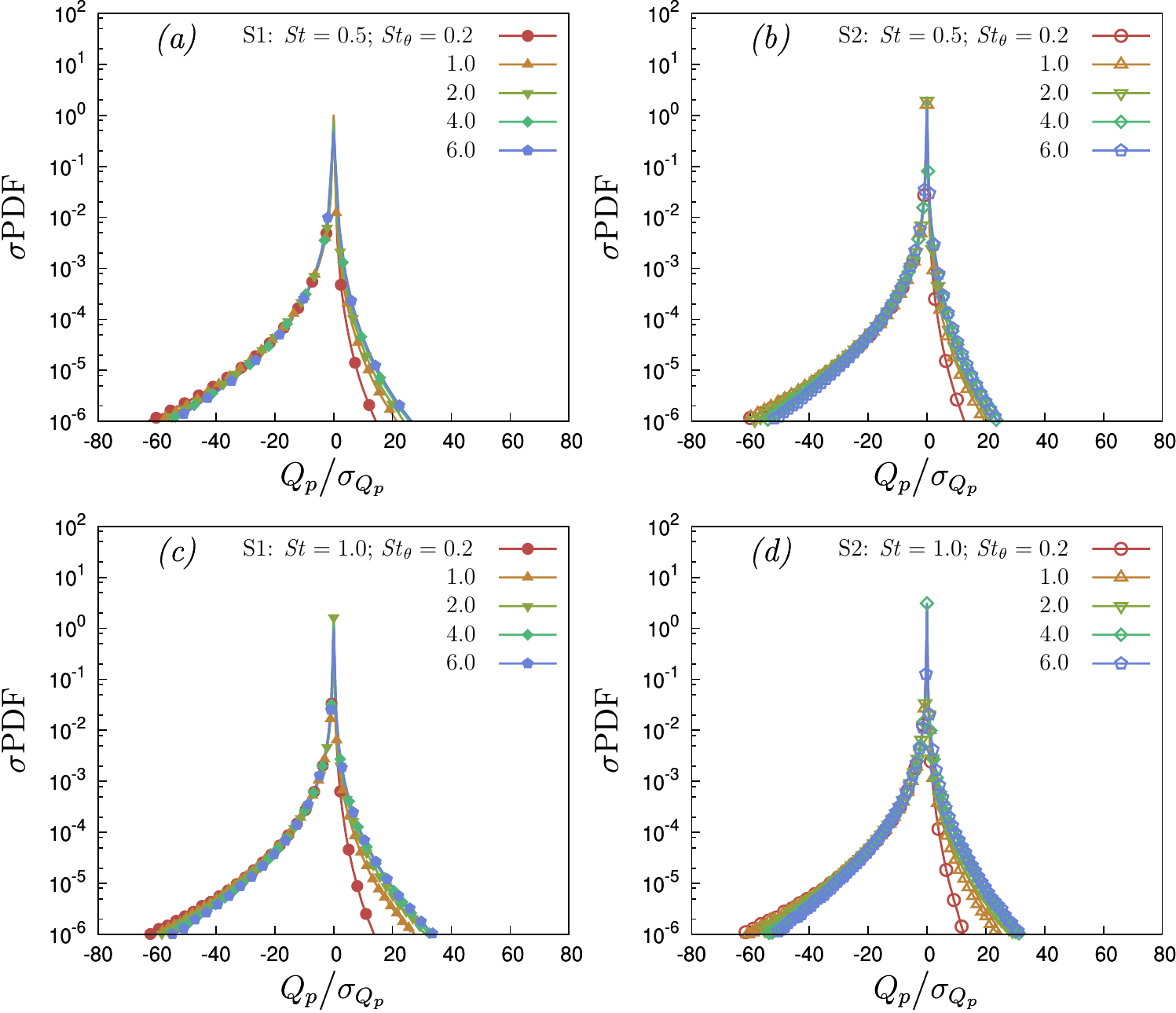}
\caption{Probability density function in normal form of the flux of particle temperature increments at small separations, $r\le2\eta$, at $\St=0.5$ (\textit{a-b}), $\St=1$ (\textit{c-d}), $\St=3$ (\textit{e-f}). The plots on the left (\textit{a-c-e}) refer to the set of simulations S1, in which the thermal particle back-reaction is included. The plots on the right (\textit{b-d-f}) refer to the set of simulations S2, in which the thermal particle back-reaction is neglected.}
\label{fig:PDFfluxdtheta}
\end{figure}

The PDF of the fluid temperature flux is plotted in normal form for $r\le2\eta$ in figure \ref{fig:PDFfluxdT}. These normalized PDFs collapse onto each other for all $\St$ and $\St_\theta$ values considered. Thus, the fluid temperature flux simply scales with its variance in the dissipation range, and the variance of the flux is modulated by the particles but the shape of the distribution is not affected by the particle dynamics. The PDF are strongly negatively skewed and have a negative mean value, associated with the mean flux of thermal fluctuations from large to small scales in the flow.

The PDF of the particle temperature flux is plotted in normal form for $r\le2\eta$ in figure \ref{fig:PDFfluxdtheta}. The PDF of the particle temperature flux across the scales is not self-similar with respect to its variance. Furthermore, the PDF becomes more symmetric as $\St_\theta$ is increased. This is associated with the increasingly non-local thermal dynamics of the particles, which allows the particle-pairs to traverse many scales of the flow with minimal changes in their temperature difference.

\section{Conclusions}
\label{sec:concl}
Using direct numerical simulations, we have investigated the interaction between the scalar temperature field and the temperature of inertial particles suspended in the fluid, with one and two-way thermal coupling, in statistically stationary, isotropic turbulence.

We found that the shape of the probability density function (PDF) of the fluid temperature gradients is not affected by the presence of the particles when two-way thermal coupling is considered, and scales with its variance. On the other hand, the variance of the fluid temperature gradients decreases with increasing $\St_\theta$, while $\St$ plays a negligible role. The PDF of the rate of change of the particle temperature, whose variance is associated with the thermal dissipation due to the particles, does not scale in a self-similar way with respect to its variance, and its kurtosis decreases with increasing  $\St_\theta$.
The particle temperature PDFs and their moments exhibit qualitatively different dependencies on $\St$ for the case with and without two-way thermal coupling.

To obtain further insight into the fluid-particle thermal coupling, we computed the number density of particles conditioned on the magnitude of the local fluid temperature. In agreement with \cite{Bec2014}, we observed that the particles cluster in the fronts of the temperature field. We also computed quantities related to moments of the particle heat flux conditioned on the magnitude of the local fluid temperature. These results showed how the particles tend to decrease the fluid temperature gradients, and that it is associated with the statistical alignments of the particle velocity and the local fluid temperature gradient field.

The two-point temperature statistics were then examined to understand the properties of the temperature fluctuations across the scales of the flow. By computing the structure functions, we observed that the fluctuations of the fluid temperature increments are monotonically suppressed as $\St_\theta$ increases in the two-way coupled regime. The structure functions of the particle temperatures revealed the dominance of thermal caustics at the small scales, wherein the particle temperature differences at small separations rapidly increase as $\St_\theta$ and $\St$ are increased. This allows particles to come into contact with very large temperature differences, which has a number of important practical implications. The scaling exponents of the inertial particle temperature structure functions in the dissipation range revealed strongly multifractal behavior. PDFs of the fluid temperature increments at different separations were found to scale in a self-similar way with their variance, just as was found for the temperature gradients. However, PDFs of the particle temperature increments do not exhibit this self-similarity, and their non-Gaussianity is much stronger than that for the fluid. 

Finally, the flux of fluid temperature increments across the scales was found to decrease monotonically with increasing $\St_\theta$. The PDFs of this flux are strongly negatively skewed and have a negative mean value, indicating that the flux is predominately from the large to the smallest scales of the flow. In the two-way coupled regime, the presence of the inertial particles does not change the shape of the PDF. The PDF of the flux of particle temperature increments in the dissipation range becomes more and more symmetric as $\St_\theta$ is increased, associated with the increasingly non-local thermal dynamics of the particles.

The results presented have revealed a number of non-trivial effects and behavior of the particle temperature statistics. In future work it will be important to consider the role of gravitation settling and coupling with water vapor fields, both of which are important for the cloud droplet problem.
Moreover, it will be interesting to include the two-way momentum coupling and to consider the non-dilute regime.

\section{Acknowledgments}
This work used the Extreme Science and Engineering Discovery Environment (XSEDE), which is supported by National Science Foundation grant number ACI-1548562 \citep{xsede}. Specifically, the Comet cluster was used under allocation CTS170009. The authors also acknowledge the computational resources provided by LaPalma Supercomputer at the Instituto de Astrofísica de Canarias through the Red Española de Supercomputación (project FI-2018-1-0044).

\bibliographystyle{jfm}
\bibliography{JFM2018}

\end{document}